\documentclass[aps,pra,reprint,twocolumn,superscriptaddress,nofootinbib, longbibliography]{revtex4-1}
\pdfoutput=1
\usepackage{graphicx}
\usepackage{bm}
\usepackage{amsmath}
\usepackage{amssymb}
\usepackage{amsfonts}
\usepackage{amsthm}
\usepackage{mathtools}
\usepackage{hyperref}
\usepackage{braket}
\usepackage[normalem]{ulem}
\usepackage{wrapfig}
\usepackage{tikz}
\usepackage{dsfont}
\usepackage{comment}
\allowdisplaybreaks

\hypersetup{colorlinks=true,urlcolor=[rgb]{0,0,0.5},citecolor=[rgb]{0.5,0,0},linkcolor=[rgb]{0,0,0.4}}

\usetikzlibrary{decorations.pathmorphing}

\newtheorem{theorem}{Theorem}

\newtheorem{corollary}{Corollary}
\newtheorem{lemma}{Lemma}

\newtheorem*{theorem*}{Theorem}
\newtheorem*{proposition*}{Proposition}

\def\autorefapp#1{\hyperref[#1]{Appendix~\ref{#1}}}

\def\pt{\widetilde{\Phi}}

\def\Real{{\mathbb R}}
\DeclareMathOperator{\Hess}{Hess}
\newcommand{\diff}{\mathop{}\!\mathrm{d}}
\def\Sphere{\mathbb{S}}
\DeclareMathOperator*{\EX}{\mathbb{E}}
\DeclareMathOperator*{\Proba}{\mathrm{Pr}}

\def\tr{{\rm tr}}

\def\and{\quad {\rm and} \quad}

\newcommand{\norm}[1]{\left\lVert#1\right\rVert}

\newcommand{\Caltech}{California Institute of Technology, Pasadena, CA 91125, USA}

\newcommand{\Berkeley}{Department of Physics, University of California, Berkeley, CA 94720, USA}

\newcommand{\MIT}{Center for Theoretical Physics, Massachusetts Institute of Technology, Cambridge, MA 02139, USA}
\newcommand{\Stanford}{Department of Mathematics, Stanford University, Stanford, CA 94305, USA}

\newcommand{\SOF}{Society of Fellows, Harvard University, Cambridge, MA 02138, USA}

\begin{document}
\title{
Emergent quantum state designs from individual many-body wavefunctions
}

\author{Jordan S.~Cotler}\thanks{These authors contributed equally to this work.}
\affiliation{\SOF}
\author{Daniel K.~Mark}\thanks{These authors contributed equally to this work.}
\affiliation{\MIT}
\author{Hsin-Yuan Huang}\thanks{These authors contributed equally to this work.}
\affiliation{\Caltech}
\author{Felipe Hern\'{a}ndez}
\affiliation{\Stanford}
\author{Joonhee Choi}
\affiliation{\Caltech}
\author{Adam L.~Shaw}
\affiliation{\Caltech}
\author{Manuel Endres}\email{mendres@caltech.edu}
\affiliation{\Caltech}
\author{Soonwon Choi}\email{soonwon@mit.edu}
\affiliation{\MIT}
\affiliation{\Berkeley}

\begin{abstract}
Quantum chaos in many-body systems provides a bridge between statistical and quantum physics with strong predictive power.  This framework is valuable for analyzing properties of complex quantum systems such as energy spectra and the dynamics of thermalization.  While contemporary methods in quantum chaos often rely on random ensembles of quantum states and Hamiltonians, this is not reflective of most real-world systems.  In this paper, we introduce a new perspective: across a wide range of examples, a single non-random quantum state is shown to encode universal and highly random quantum state ensembles.  We characterize these ensembles using the notion of quantum state $k$-designs from quantum information theory and investigate their universality using a combination of analytic and numerical techniques.  In particular, we establish that $k$-designs arise naturally from generic states as well as individual states associated with strongly interacting, time-independent Hamiltonian dynamics.  Our results offer a new approach for studying quantum chaos and provide a practical method for sampling approximately uniformly random states; the latter has wide-ranging applications in quantum information science from tomography to benchmarking.
\end{abstract}

\maketitle

Analyzing the exact dynamics of general, strongly interacting quantum many-body systems is intractable using existing analytic and numerical tools.
However, there is a widely used heuristic for understanding chaotic quantum dynamics: the eigenstates and eigenvalues of chaotic Hamiltonians have properties as if they were sampled from a random ensemble.
This heuristic leads one to leverage statistical approaches, such as random matrix theory~\cite{wigner1993characteristicI, dyson1962statisticalI}, to address many physical problems and has become a foundational principle in understanding chaos and thermalization in quantum systems~\cite{popescu2006entanglement, reimann2008foundation, linden2009quantum, short2012quantum}.  Examples of this heuristic include Berry's conjecture~\cite{Berry1977} and the eigenstate thermalization hypothesis (ETH)~\cite{Deutsch1991Quantum,Srednicki1994,Rigol2008}, which have been supported by an overwhelming amount of numerical evidence~\cite{Nandkishore2015,Alessio2016,abanin2019colloquium}.

While these statistical approaches hinge on the presence of random ensembles, realistic quantum systems are often described by a \emph{fixed} Hamiltonian.
Accordingly, energy eigenstates, eigenvalues, and states evolved by Hamiltonian dynamics
are deterministic functions of a particular Hamiltonian without any randomness involved.
Thus, it is of fundamental interest to
establish a connection between isolated quantum systems and the emergence of random ensembles that dictate their statistical properties.

In this paper, we present a new perspective on the emergence of statistical behavior in chaotic quantum many-body systems: instead of imagining that a physical state is sampled from a random ensemble, we use a single many-body wavefunction to \textit{generate} an ensemble of pure states on a subsystem; this ensemble goes beyond the reduced density matrix.
Then we leverage statistical tools to analyze the degree of randomness of the local ensemble, and in turn characterize the global wavefunction.

\begin{figure*}[t!]
	\centering	\includegraphics[width=145mm]{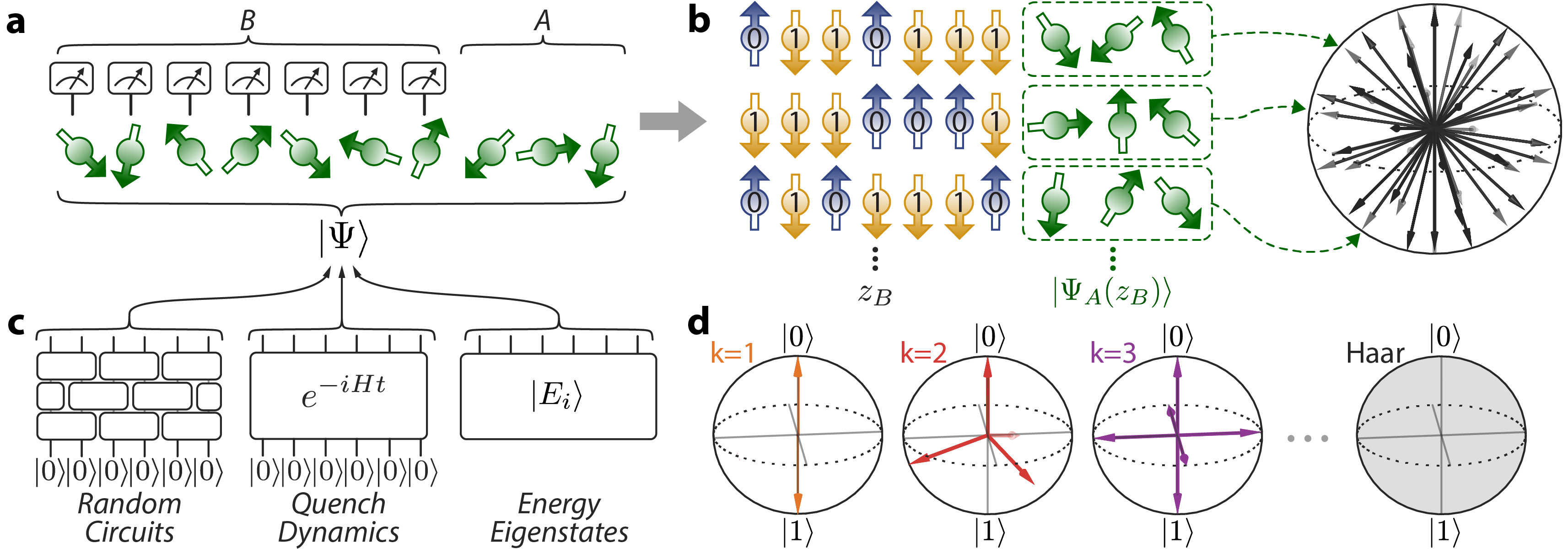}
	\caption{
	\textbf{Emergence of a universal quantum state ensemble from a single many-body wavefunction.}
	\textbf{a,} A subsystem $B$ of a pure many-body wavefunction $\ket{\Psi}$ is measured in a fixed local basis;
	the remaining unmeasured qubits in $A$ are in a pure state that depends on the measurement outcome on $B$.
	\textbf{b,} For quantum systems consisting of qubits, the measurement on $B$ samples random outcomes, each characterized as a bitstring $z_B$ (binary numbers in the blue/yellow arrows).
	Different measurement outcomes $z_B$ occur with probability $p(z_B)$ and lead to distinct quantum states $\ket{\Psi_A(z_B)}$, forming the projected ensemble $\mathcal{E} = \{p(z_B), |\Psi_A(z_B)\rangle\}$. Right panel: the ensemble of pure states $\ket{\Psi_A(z_B)}$ (black arrows) are randomly distributed in the Hilbert space of $A$ (black sphere), forming an approximate quantum state $k$-design.
	\textbf{c,}
	Examples of many-body wavefunctions whose projected ensembles form approximate quantum state designs include 
	the typical output states of random unitary circuit evolution, 
	quantum states obtained from quenched time evolution,
	and energy eigenstates of a chaotic Hamiltonian at infinite temperature.
	\textbf{d,} Illustration of minimal quantum state $k$-designs for a single qubit on the Bloch sphere. 
	Forming higher $k$-designs requires a larger number of pair-wise non-orthogonal quantum states. In the limit of $k \rightarrow \infty$, a $k$-design approaches the so-called Haar ensemble, which is the uniform distribution over all pure quantum states.
	 } \vspace{-0.2cm}
	\label{Fig1}
\end{figure*}

Concretely, an ensemble of states can be generated from a single wavefunction by performing local measurements over only part of the total system.
We consider a many-body system partitioned into a subsystem $A$ and its complement $B$. Performing local measurements on $B$, we obtain exponentially many different pure states on $A$, each corresponding to a distinct measurement outcome on $B$.
We call the set of pure states on $A$, along with the associated measurement probabilities,
the \textit{projected ensemble} on $A$ (see Fig.~\ref{Fig1}a,b); see also~\cite{Choi2021}.
To quantify the degree of randomness of an ensemble,
we use a well-established notion from quantum information theory, namely \emph{quantum state $k$-designs}~\cite{renes2004symmetric, ambainis2007quantum}.
An ensemble of pure states is said to form a quantum state $k$-design if the probability distribution over the constituent states is indistinguishable up to $k$-th moments from the uniform distribution over the entire Hilbert space.
Therefore, forming a higher $k$-design implies that the ensemble is more uniformly distributed over the Hilbert space.

Remarkably, using our new approach, we find that approximate $k$-designs arise from a variety of natural quantum many-body states. 
We establish two theorems showing that the projected ensemble coming from generic many-body quantum states forms an approximate $k$-design as long as the size of $B$ is sufficiently larger than the size of $A$.
Furthermore, a concurrent work~\cite{Choi2021} finds evidence that approximate $k$-designs emerge from projected ensembles in a Rydberg quantum simulator.
We argue that this is a much more general phenomenon: we find strong numerical evidence across several models that approximate $k$-designs arise from both quantum states obtained by quenched time evolution and energy eigenstates of chaotic Hamiltonians (see Fig.~1c).
In the former case, we find that the degree of randomness in a projected ensemble continues to grow even after conventional local thermalization has occurred.  Specifically, we establish numerically that 
states evolved for a longer amount of time form higher approximate $k$-designs.
In the case of energy eigenstates, approximate $k$-designs emerge from states in the middle of the energy spectrum corresponding to effective infinite temperature.
For finite-temperature eigenstates we observe that the projected ensembles converge to a universal ensemble that smoothly varies with respect to the energy density.

Our findings suggest that for a wide range of physically relevant many-body states, projected ensembles exhibit a universal form of randomness.
This allows us to quantify the chaotic nature of Hamiltonian dynamics and to study the growth of complex, nonlocal correlations between a subsystem and its complement beyond the conventional paradigm of quantum thermalization~\cite{Deutsch1991Quantum,Srednicki1994,Rigol2008,Nandkishore2015,kaufman2016quantum,abanin2019colloquium,ueda2020quantum}.  Post-thermalization physics in quantum many-body systems exhibits interesting quantitative and qualitative differences from its classical counterpart (e.g.~\cite{Cotler2020}), and the current work clarifies and extends this paradigm.  Furthermore, our work presents a simple protocol to efficiently produce an ensemble of random pure states from fixed, time-independent Hamiltonian dynamics that does not require highly fine-tuned controls.

\section{Quantum state designs from projected ensembles}
\label{Sec:maintheorems}
Studying properties of random quantum states is useful because they often encode universal phenomena found in nature.  A standard approach is to consider the \emph{Haar ensemble}, which is the uniform distribution over all pure states in a Hilbert space.  While the Haar ensemble can be studied analytically using various statistical tools~\cite{harrow2013church}, it is extremely challenging to experimentally realize, as doing so requires an exponential amount of resources (such as the number of quantum gates or experimental operations)~\cite{knill1995approximation}.

Instead, ensembles that form quantum state designs are considered because they mimic the Haar ensemble and can be efficiently realized in physical systems~\cite{brandao2016local, Harrow2009, nakata2017efficient, harrow2018approximate, haferkamp2020quantum,farshi2020time}.
To build intuition for quantum state designs, let us consider quantum state designs for a single qubit (see Fig.~1d).
For example, the ensemble of single-qubit states $\{|0\rangle, |1\rangle\}$
with equal probabilities has the same mean (first moment) as
the Haar ensemble: $\rho^{(1)} = \frac{1}{2} \left(|0\rangle \langle 0| + |1 \rangle \langle 1|\right) = \EX_{\Psi\sim \textrm{Haar}}\!\left[ |\Psi\rangle \langle \Psi|\right]$,
where $\EX_{\Psi \sim \mathcal{E}}$ denotes averaging $\ket{\Psi}$ over the ensemble $\mathcal{E}$.
We say that such a two-state ensemble forms a $1$-design.
However, the two-state ensemble does not form a $2$-design because its second moment $\frac{1}{2}\left(|0\rangle \langle 0|\otimes |0\rangle \langle 0| + |1\rangle \langle 1|\otimes |1\rangle \langle 1|\right)$ differs from that of the Haar ensemble.
To form a 2-design for a single qubit, one can use four distinct non-orthogonal quantum states uniformly spread over the Bloch sphere (Fig.~1d).
In general, for an ensemble to form a higher order design, it must be supported over a larger number of states~\cite{delsarte1991spherical}.

Formally, given an ensemble $\mathcal{E} = \{p_i, |\Psi_i\rangle \}$ consisting of a set of wavefunctions $\ket{\Psi_i}$ and their probabilities $p_i$, one can test whether the ensemble forms an approximate $k$-design by explicitly comparing its $k$-th moment
\begin{align}
\label{Eq:momentState}
\rho^{(k)}_\mathcal{E} = \EX_{\Psi \sim \mathcal{E}}\!\left[ \left( |\Psi\rangle \langle \Psi|\right)^{\otimes k} \right] = \sum_i p_i\left(|\Psi_i\rangle \langle \Psi_i|\right)^{\otimes k}
\end{align}
against that of the Haar ensemble $\rho_\textrm{Haar}^{(k)}$ (which is defined by a continuous probability distribution over pure states).
More precisely, we say an ensemble $\mathcal{E}$ is an $\varepsilon$-approximate quantum state $k$-design if
\begin{align}
\label{Eq:1-norm}
&\left\| \rho^{(k)}_\mathcal{E} - \rho^{(k)}_\textrm{Haar}  \right\|_1 \leq \varepsilon\,,
\end{align}
where $\| \cdot \|_1$ denotes the trace norm.
This definition means that the $k$-th moment of $\mathcal{E}$ is nearly indistinguishable from the $k$-th moment of the Haar ensemble up to a small error $\varepsilon$.  It can be shown that an $\varepsilon$-approximate $k$-design is also an $\varepsilon$-approximate $j$-design for any $j<k$, and accordingly larger values of $k$ indicate that an ensemble looks more uniformly random.

Unlike the Haar ensemble, approximate designs arise in physical settings~\cite{Harrow2009, brandao2016local, nakata2017efficient, harrow2018approximate, haferkamp2020quantum, farshi2020time}.
A canonical example is random unitary circuits \cite{Harrow2009, brandao2016local, harrow2018approximate}, where a set of random two-qubit unitary gates in a certain geometric arrangement are sequentially applied to simple initial states (for an example, see Fig.~1c).
Then the ensemble of resulting states over different choices of unitary gates forms an $\varepsilon$-approximate $k$-design as long as the depth of the circuit is sufficiently large, scaling polynomially in $k$, $\log(1 / \varepsilon)$, and the number of qubits $N$~\cite{brandao2016local}.
Similarly, there are a number of proposals to generate approximate designs based on time-dependent local Hamiltonian evolution~\cite{nakata2017efficient, farshi2020time}.

In these examples, states are sampled from approximate $k$-designs by realizing many distinct, highly engineered quantum evolutions.
By contrast, we show that approximate $k$-designs arise naturally from the projected ensemble of a \textit{single} many-body wavefunction.  More precisely, consider a many-body wavefunction $\ket{\Psi}$ in the Hilbert space $\mathcal{H}$ for a bipartite system consisting of $N_A$ qubits in $A$ and $N_B$ qubits in $B$.
The projected ensemble for $A$ is generated by performing projective measurements on all $N_B$ qubits in $B$ in a local basis $\{\ket{z_B}\}$, where a bitstring $z_B \in \{0,1\}^{N_B}$ enumerates over all $2^{N_B}$ measurement outcomes.
After the measurement, with probability $p(z_B)$ the system is described by the normalized wavefunction $\ket{\Psi_A(z_B)}\otimes \ket{z_B}$, where
\begin{align}
p(z_B) &:= \bra{\Psi} \left( \mathds{1}_A \otimes \ket{z_B}\bra{z_B} \right) \ket{\Psi}\\
|\Psi_A(z_B)\rangle &:= \big(\mathds{1}_A \otimes \langle z_B| \big) |\Psi\rangle / \sqrt{p(z_B)}\,
\end{align}
and $\mathds{1}_A$ is the identity operator acting on qubits in $A$.
This defines a \textit{projected ensemble}:
\begin{equation}
\label{eq:proj_ensemble}
\mathcal{E}_{\Psi, A} := \{p(z_B)\,,\,|\Psi_A(z_B)\rangle\}\,.
\end{equation}
The ensemble consists of $2^{N_B}$ states which are generally not pairwise orthogonal.
The projected ensemble contains strictly more information than the reduced density matrix,
since the density matrix is the first moment of the ensemble.
We will call $|\Psi\rangle$ the \textit{generator state} of $\mathcal{E}_{\Psi,A}$.
We note that Refs~\cite{verstraete2004entanglement, popp2005localizable} use a related construction to define localizable entanglement.

\begin{figure*}[t]
	\centering
	\includegraphics[width=1\textwidth]{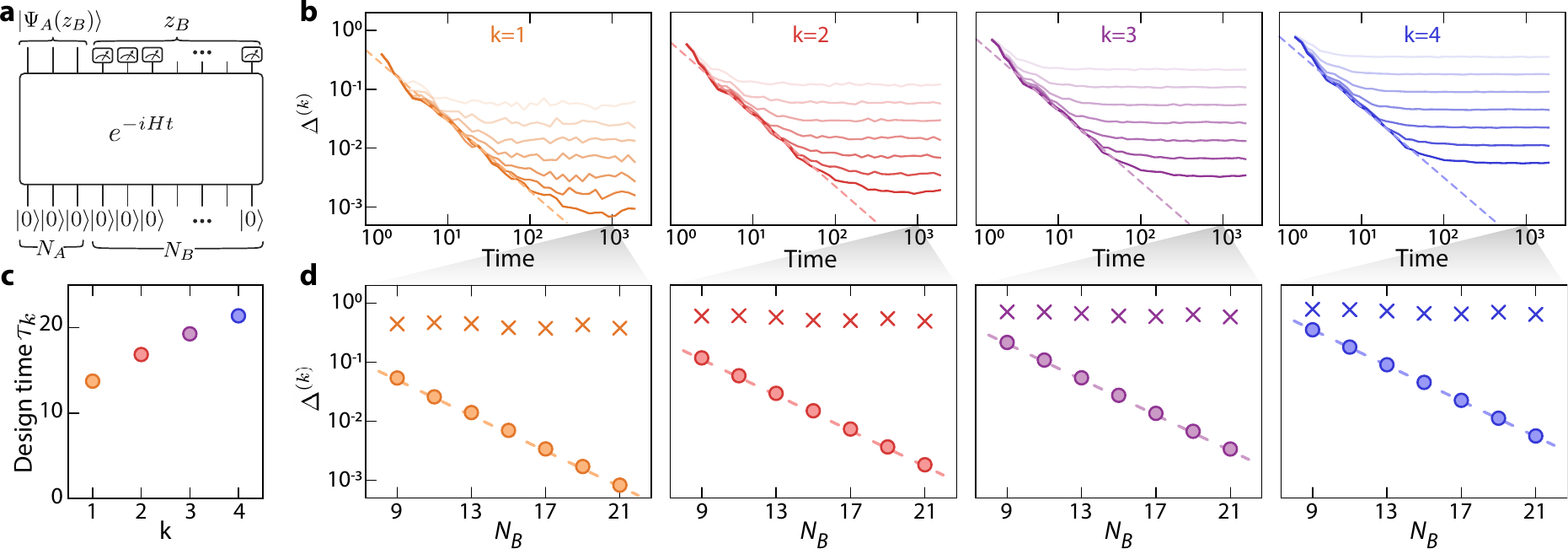}
	\caption{
	\textbf{Emergent quantum state designs from chaotic time evolution.}
	\textbf{a,} Quenched dynamics under a time-independent Hamiltonian $H$ starting from an initial product state, $\ket{0}^{\otimes N}$, for a $N$-qubit system. The system is partitioned into two subsystems $A$ and $B$ with size $N_A$ and $N_B$, respectively. At time $t$, a projective measurement in the local $z$-basis is performed on subsystem $B$, resulting in a specific outcome $z_B$ of length $N_B$.
	\textbf{b,} Trace distances $\Delta^{(k)}$ between the $k$-th moments of the Haar ensemble and projected ensembles for an $N_A=3$ subsystem as a function of evolution time for various total system sizes $N=12, 14,\dots 24$ (darker colors for increasing $N$).
	Dashed lines are a phenomenological power-law fit, yielding a scaling of $\Delta^{(k)} \sim t^{-1.2}$.
	 \textbf{c,} Design time $\tau_k$ defined by the evolution time to achieve a trace distance of $\Delta^{(k)} = 0.02$. We find that longer time evolution is required to form a higher approximate $k$-design.
	  \textbf{d,} Late-time trace distances at $t\approx 10^3$. For a mixed-field chaotic Hamiltonian (circles), the late-time trace distances exhibit an exponential scaling with $N_B$ while they remain nearly constant for an integrable, non-ergodic Hamiltonian (crosses, time traces not shown). Dashed lines represent the trace distances $\Delta^{(k)}_\text{em}$ between the $k$-th moments of the Haar ensemble and a finite set of $2^{N_B}$ states sampled from the Haar ensemble on $A$.
	} \vspace{-0.5cm}
	\label{Fig2}
\end{figure*}

We find that projected ensembles form approximate $k$-designs for many physically relevant many-body wavefunctions, including special, analytically soluble cases.
Specifically, it can be shown that a generator state for an $\varepsilon$-approximate $k$-design can be efficiently prepared by a simple protocol, in which an initial product state is evolved by a Hamiltonian over a short time duration.
The Hamiltonian can be time-independent and only has nearest-neighbor Ising interactions.
This is possible because a universal resource state for measurement-based quantum computation can be prepared within a constant time independent of system size~\cite{Raussendorf2001OnewayQC,Mezher2018EfficientPseudorandomness}.  While this example is fine-tuned, we show that, in fact, \textit{generic} many-body wavefunctions are good generator states for $\varepsilon$-approximate $k$-designs:
\begin{theorem}\label{Thm:HaarThm} Let $|\Psi\rangle$ be chosen uniformly at random from the Hilbert space $\mathcal{H}$.
The ensemble $\mathcal{E}_{A, \Psi}$ forms an $\varepsilon$-approximate $k$-design with probability at least $1-\delta$ if
\begin{equation}
N_B = \Omega\left(k\, N_A + \log\left( \tfrac{1}{\varepsilon} \right) + \log\log\left(\tfrac{1}{\delta}\right)\right).
\end{equation}
\end{theorem}
\noindent
Here
$\Omega(\cdot)$ denotes a lower bound up to a constant multiplicative factor and subleading terms.
This theorem is proved in Appendix~\ref{sec:proofsthm}, and establishes that all but a tiny fraction (of order $\sim 1/2^{2^{N_B}}$) of the states in the Hilbert space are generator states for approximate $k$-designs if $N_B$ is asymptotically larger than $k$ times $N_A$.
However, a quantum state randomly sampled from the entire Hilbert space is not so physical since such a state is extremely difficult to produce experimentally~\cite{knill1995approximation}.
To this end, we also present the following theorem:
\begin{theorem}
\label{Thm:tdesigngen}
Let $\ket{\Psi}$ be a state sampled from an ensemble on $\mathcal{H}$ that forms an $\varepsilon'$-approximate $k'$-design. Then the projected ensemble $\mathcal{E}_{A, \Psi}$ forms an $\varepsilon$-approximate $k$-design with probability at least $1-\delta$ if
\begin{align}
N_B &= \Omega\left( k N_A + \log\left( \tfrac{1}{\varepsilon \delta} \right) \right),\\
k' &= \Omega\left( k \left(N_B +\log\left(\tfrac{1}{\varepsilon \delta}\right) \right)\right),\\
\log\left(\tfrac{1}{\varepsilon'}\right) &= \Omega\left( k N_B \left(N_B + \log\left(\tfrac{1}{\varepsilon \delta}\right)\right) \right),\\
N_A &= \Omega\left( \log(N_B) +\log(k) + \log\log\left(\tfrac{1}{\varepsilon \delta}\right) \right)\,.
\end{align}
\end{theorem}
\noindent The proof is given in Appendix~\ref{sec:proofsthm} and relies on higher-order concentration of measure results~\cite{bobkov2019higher} as well as a polynomial approximation technique used in quantum algorithms for solving linear systems~\cite{childs2017quantum}.
This theorem shows that if the generator state is complex enough, in the sense that it is a typical state from an $\varepsilon'$-approximate $k'$-design for small $\varepsilon'$ and large $k'$~\cite{brandao2019models}, then the projected ensemble will well-approximate a quantum $k$-design.

Theorem~\ref{Thm:tdesigngen} has implications for ongoing experiments: if a \textit{single sample} of an approximate design is experimentally realized, our theorem states that it can be used to generate \textit{ensembles} forming approximate designs on its subsystems.  Moreover, this protocol is hardware-efficient since it does not require fine-tuned controls to produce many different quantum states, and could lead to various useful applications in quantum information science~\cite{brandao2016local, alagic2018unforgeable, dankert2009exact, cross2019validating, elben2020cross, huang2020predicting}.

At a conceptual level, Theorem~\ref{Thm:tdesigngen} establishes that a large class of states which can be efficiently prepared are good generators of $k$-designs. 
This raises the possibility that, even in natural chaotic quantum systems, approximate $k$-designs may arise from projected ensembles.

\section{Quantum State Designs from Chaotic Systems}
\label{Sec:numerics}
Motivated by the above
results, we numerically investigate projected ensembles that arise from chaotic Hamiltonian dynamics. 
We start with the paradigmatic example of the 1D quantum Ising spin system with mixed fields (QIMF), described by the Hamiltonian
\begin{align}
H_\textrm{QIMF} = h^x \sum_{j=1}^N \sigma_j^x + h^y \sum_{j=1}^N \sigma_j^y + J \sum_{j=1}^{N-1} \sigma_j^x \sigma_{j+1}^x\,,
\end{align}
where $N$ is the number of spins,  $\sigma^{\mu}_j$ with $\mu=x,y,z$ are the Pauli operators for a spin at site $j$, $J$ is the strength of Ising interactions, and $h^x$ and $h^y$ are the strengths of the longitudinal and transverse fields, respectively. In the absence of the longitudinal field ($h^x = 0$), the Hamiltonian can be mapped to an integrable model of non-interacting fermions via the Jordan-Wigner transformation, leading to non-ergodic dynamics. However, for any non-zero longitudinal field ($h^x \neq 0$), the Hamiltonian is ergodic, and its eigenvalues and eigenvectors are expected to have properties consistent with ETH predictions.
This has been explicitly checked for a specific parameter set $(h^x, h^y,J) = (0.8090, 0.9045, 1)$~\cite{kim2014testing} which we adopt for our study.
We note that we do not find a qualitative differences in our results when using nearby parameter values.

We first consider the many-body state $\ket{\Psi(t)} = e^{-i H_\textrm{QIMF} t} \ket{\Psi_0} $ resulting from time evolution of the initial state $\ket{\Psi_0} = \ket{0}^{\otimes N}$ (Fig.~2a). Here $\ket{0}_j$ and $\ket{1}_j$ are the eigenstates of $\sigma_j^z$ with eigenvalues $+1$ and $-1$, respectively. The initial product state, $\ket{\Psi_0}$, has zero energy expectation value with respect to $H_\textrm{QIMF}$, corresponding to the total energy of an infinite temperature state. This implies that local subsystems will relax to an infinite-temperature ensemble after a local thermalization time~\cite{Deutsch1991Quantum,Srednicki1994,Rigol2008}.

At any time $t$, the projected ensemble for a subsystem $A$ is obtained by simulating projective measurements on the rest of the $N_B=N-N_A$ qubits in the local $z$-basis (Fig.~2a).\footnote{We note that our measurement $z$-basis is orthogonal to the Hamiltonian; this choice is made to ensure that the measurement outcomes are not explicitly correlated with the total energy in subsystem $A$.}  In order to check if the projected ensemble forms an approximate $k$-design, we compare the $k$-th moment of the ensemble, $\rho^{(k)}_{\mathcal{E}}$ in Eq.~(1), to that of the uniform ensemble $\rho^{(k)}_\textrm{Haar}$ using the trace distance $\Delta^{(k)} = \frac{1}{2} \left\| \rho^{(k)}_{\mathcal{E}} - \rho^{(k)}_\textrm{Haar} \right\|_1$. For a fixed subsystem size of $N_A=3$, we numerically compute the trace distance $\Delta^{(k)}$ up to $k=4$ as a function of time for various $N_B$ (Fig.~2b).
In all cases, $\Delta^{(k)}$ decays in time, following a phenomenological power-law scaling $\Delta^{(k)} \sim 1/t^{1.2}$, until it saturates to a value that is governed by finite size effects. We note that the extracted scaling exponent is non-universal and Hamiltonian-dependent.
The saturation value of $\Delta^{(k)}$ decreases exponentially with $N_B$ (circles, Fig.~2d), exhibiting the scaling $\Delta^{(k)} \sim 1/\sqrt{2^{N_B}}$.
If the power-law scaling persists at larger system sizes then $\Delta^{(k)}$ may decrease over an exponentially long time scale.

As a comparison, we also present $\Delta^{(k)}_\text{em}$ which is the trace distance between the $k$-th moments of the Haar ensemble and the empirical Haar ensemble consisting of $2^{N_B}$ states sampled uniformly at random on $A$ (dashed lines, Fig.~2d).  Since the empirical ensemble asymptotically approaches the Haar ensemble in the limit of infinite samples, $\Delta^{(k)}_\text{em}$ is determined only by statistical fluctuations associated with having a finite number of quantum states.
Remarkably, we find that the projected ensemble obtained from the quench dynamics shows a trace distance almost identical to that of the
empirical ensemble of the same size, suggesting that the former is as uniformly random as the latter.
By contrast, repeating similar calculations for the integrable model ($h^x=0$), we observe qualitatively different behavior where the trace distance to the Haar ensemble is much larger than in the non-integrable case (crosses, Fig.~2d).  Furthermore, there is no appreciable dependence on system size.  This is expected, since integrable systems do not locally thermalize and instead relax to a generalized Gibbs ensemble, and hence will not form 1-designs at effective infinite temperature~\cite{vidmar2016generalized}.

Given the emergence of $k$-designs in asymptotic regimes for a chaotic Hamiltonian, it is natural to ask how long it takes for a subsystem to achieve an approximate design up to a small, fixed precision.
To this end, we introduce a design time $\tau_k$ defined as the time at which $\Delta^{(k)}$ becomes smaller than a certain fixed threshold $\varepsilon$. For a chosen threshold $\varepsilon = 0.02$, we find that the formation of higher $k$-designs requires longer time evolution (Fig.~2c).
This observation is consistent with the idea that typical quantum states from higher $k$-designs are more complex and hence more difficult to prepare~\cite{roberts2017chaos, cotler2017chaos, brandao2019models}.

\begin{figure}[t!]
	\centering
	\includegraphics[width=\columnwidth]{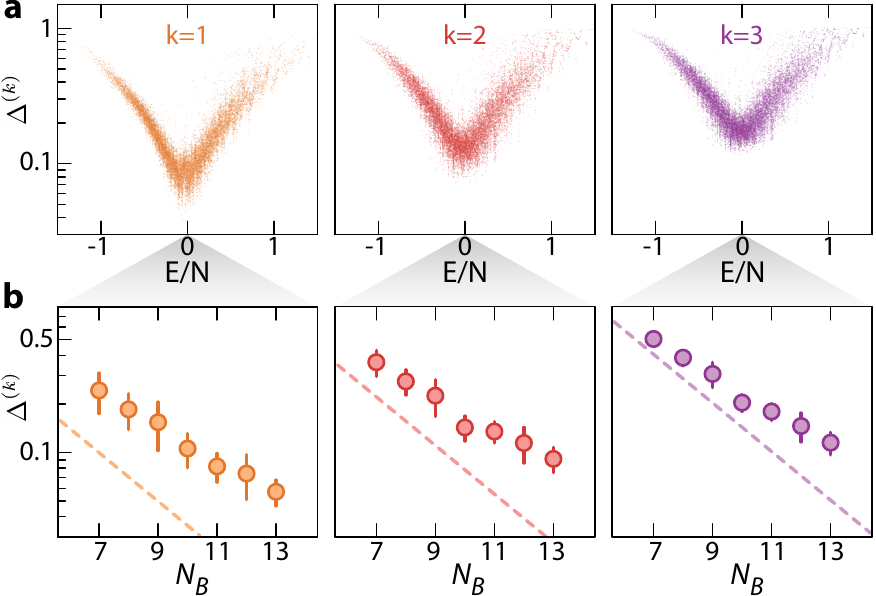}
	\caption{
	\textbf{Emergent quantum state designs from energy eigenstates.}
	\textbf{a,} Trace distances between the $k$-th moments of the Haar ensemble and a projected ensemble for an $N_A=3$ subsystem generated from the energy eigenstates of a Hamiltonian. Results are presented as a function of energy density $E/N$ for a total system size of $N=14$.
	\textbf{b,} Trace distances for the projected ensembles obtained from eigenstates near zero energy corresponding to infinite temperature. The distances exhibit an exponential decay as a function of system size. The points are evaluated for 100 eigenstates near zero energy, and the error bars denote their standard deviation. For comparison, dashed lines represent the trace distances $\Delta^{(k)}_\text{em}$ between the Haar ensemble and an empirical ensemble of $2^{N_B}$ states sampled from the Haar ensemble.
	} \vspace{-0.5cm}
	\label{Fig3}
\end{figure}

Next we investigate the properties of energy eigenstates. We repeat a similar analysis as above by replacing the time-evolved state $\ket{\Psi(t)}$ with an energy eigenstate $\ket{E_i}$ of $H_\textrm{QIMF}$. Figure~3 shows the trace distance $\Delta^{(k)}$  as a function of energy $E_i$ for a projected ensemble generated from $|E_i\rangle$.
We find a sharp dip at zero energy density corresponding to infinite temperature, which signifies the emergence of approximate quantum state designs (Fig.~3a).
At zero energy, $\Delta^{(k)}$ decreases exponentially as a function of system size for $k=1,2,3$.  However, the decay rate is slightly slower than in the case of quenched dynamics (Fig.~3b).

In addition to the QIMF, we have also studied two other ergodic Hamiltonian models in order to corroborate the universality of our findings (see Appendix~\ref{sec:morenumerics}).
Specifically, we consider a system of random, all-to-all coupled spin-1/2 particles as well as hard-core bosons with random all-to-all hoppings with particle number conservation.
In both cases, we observe excellent convergence of projected ensembles to approximate $k$-designs. 
In the latter case, the measurement outcomes in $B$ and corresponding pure quantum states $A$ are strongly correlated owing to the particle number conservation symmetry; hence  a na\"{i}ve approach based on Eq.~\eqref{eq:proj_ensemble} does not lead to approximate $k$-designs. We instead introduce symmetry-resolved projected ensembles by grouping certain subsets of measurement outcomes from $B$ (see Appendix~\ref{sec:morenumerics} for details); this does lead to approximate $k$-designs.

For chaotic Hamiltonians, projected ensembles forming $k=1$ designs can be anticipated from the standard picture of quantum thermalization since the first moment of a projected ensemble simply corresponds to the reduced density matrix of a subsystem.
The reduced density matrix approaching the first moment of the Haar ensemble, i.e.~the maximally mixed state, follows from local thermalization at infinite temperature.
However, the convergence of higher moments $k \geq 2$ of the projected ensemble to higher $k$-designs is nontrivial and surprising.
Such convergence cannot be explained by ETH alone, and suggests a new form of emergent randomness beyond the conventional framework of quantum thermalization.

\begin{figure}[t!]
	\centering
	\includegraphics[width=\columnwidth]{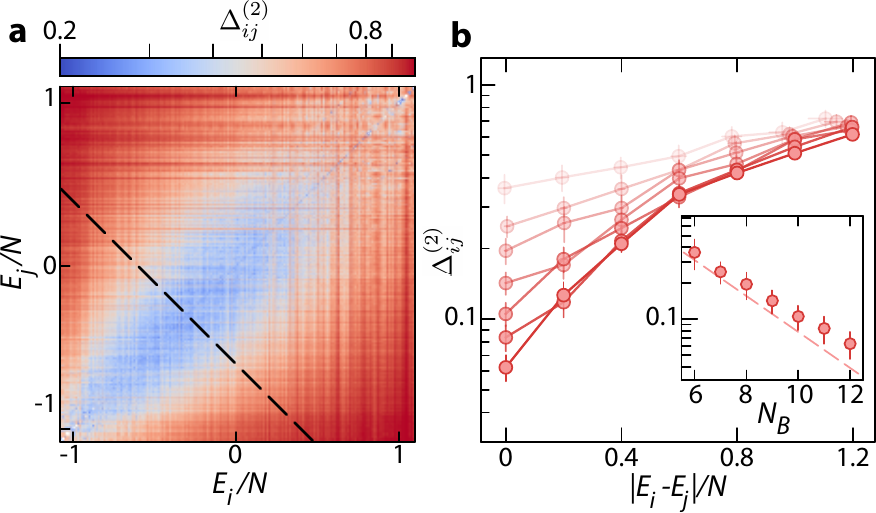}
	\caption{
	\textbf{Universal ensemble at finite temperature.}
	\textbf{a,} Trace distances between the second moments of projected ensembles generated from a pair of energy eigenstates at $E_i$ and $E_j$ for $N_A = 3$ subsystems. We plot the pairwise distances $\Delta^{(2)}_{ij}$ for every pair of eigenstates $|E_i\rangle, |E_j\rangle$, computed for system size $N=11$. The distances are minimized when $E_i \approx E_j$. The plot suggests the existence of a universal ensemble that depends smoothly on the energy density.
	The black dashed line indicates a cut defined by $(E_i+ E_j)/N = -0.6$.
	\textbf{b,} Trace distances plotted as a function of energy density difference, $|E_i - E_j|/N$, along the black dashed line in \textbf{a} for various systems sizes $N=9, 10, \dots, 15$ (darker colors for increasing $N$).
	Inset: the distances at zero energy difference $E_i = E_j$ decay exponentially with system size. The trace distances from the projected ensemble are comparable to those from the finite-size ensemble, $\Delta^{(2)}_\text{em}$\,, of the empirical Haar distribution (dashed line). Such an exponential scaling suggests the existence of a universal random ensemble at {\it finite} temperatures.
	} \vspace{-0.5cm}
	\label{Fig4}
\end{figure}

A natural next step is to generalize quantum state designs (or Haar ensembles) to a finite-temperature setting.
However, we are unaware of any appropriate analogue of the Haar ensemble at finite temperature.
Such an ensemble, if it exists, would generally depend on the system Hamiltonian, and its first moment should (approximately) be a thermal state.  While explicitly identifying properties of such an ensemble is an interesting future direction, here we find numerical evidence that such an ensemble exists.
In Figure 4, we compute projected ensembles for all energy eigenstates of $H_\textrm{QIMF}$  and present the pairwise distances
\begin{align}
\Delta_{ij}^{(2)} = \frac{1}{2}\left\|\rho_{i}^{(2)} - \rho_{j}^{(2)} \right\|_1\,,
\end{align}
where $\rho_{i}^{(2)}$ denotes the second moment of a projected ensemble generated from an eigenstate $\ket{E_i}$.
We find that $\Delta_{ij}^{(2)}$ is a smooth function of energy up to small fluctuations, suggesting that the projected ensembles smoothly vary with energy as well (Fig.~4a).
Also, $\Delta_{ij}^{(2)}$ is minimized when the energy difference $|E_i - E_j|$ is small (Fig.~4b), and in this regime $\Delta_{ij}^{(2)}$ decreases exponentially with system size (Fig.~4b inset).
These observations suggest that the second moment of the projected ensemble is indeed universal even at finite temperatures.

\section{Discussion and Outlook}
\label{Sec:discussion}
The random quantum state ensembles considered in this paper are qualitatively different from conventional ones in quantum statistical mechanics, such as the microcanonical and canonical ensembles.  These latter ensembles are fully specified by their corresponding density matrices (their first moment) and are used to evaluate expectation values of observables.

Our formalism concerns more general statistical properties (such as higher moments) of an ensemble with a large number of pure states that are generally pairwise non-orthogonal.
As such, a projected ensemble encodes additional information about a subsystem.
For example, one can ask the following information-theoretic question: how much information (i.e., classical bits) is required in order to specify the full wavefunction of a random sample from a projected ensemble?
If the projected ensemble were to be the Haar ensemble, an exponential number of bits would be required.
By contrast, for an ensemble which is uniformly distributed only over computational basis states of $A$, a linear number of bits suffice to specify a sample.
Notice that the former and latter ensembles produce the same density matrix, namely the maximally mixed state.  Therefore, projected ensembles provide a novel framework to analyze the information content associated with quantum states of a subsystem and their relation with the remainder of the system.

Our findings open up a number of new directions in quantum chaos, thermalization, and quantum information. In particular, our numerical results demonstrate that for an initial product state evolved by a chaotic Hamiltonian, the largest $k$-design attained by the projected ensemble grows as a function of time; moreover, this growth persists significantly past the local thermalization timescale.  While it is not presently clear how long this growth will persist, the observed growth is clearly diagnostic of the sustained development of non-local correlations after local thermalization has occurred (e.g.~\cite{cotler2020spectral, Cotler2020}).  There may even be connections to the quantum complexity of a state evolving by chaotic dynamics, which likewise grows long after thermalization has occurred~\cite{brown2018second, brandao2019models}.  It would also be interesting to fully generalize the above to projected ensembles at finite temperature, in the presence of symmetries, (quasi-)integrability, or strong disorder resulting in localization~\cite{Nandkishore2015,Alessio2016,abanin2019colloquium}. 

In quantum information science, quantum states designs are valuable resources in many applications. Our work establishes projected ensembles as a practical way of sampling states from approximate designs using natural Hamiltonian evolutions of existing quantum simulators without fine control. Further, our work could lead to novel experimental quantum tomography protocols~\cite{huang2020predicting}, cryptographic protocols for hiding information~\cite{brandao2016local}, the design of unforgeable quantum encryption~\cite{alagic2018unforgeable}, and also new methods for quantum device verification~\cite{elben2020cross,carrasco2021theoretical}.  Indeed, a parallel work~\cite{Choi2021} uses projected ensembles to devise and implement a novel benchmarking protocol.

Finally, an important question at the intersection of computer science and  quantum many-body physics is whether the computational complexity of simulating natural chaotic dynamics is beyond the capability of contemporary classical computers. In other words, can quantum supremacy tests be performed using a fixed chaotic Hamiltonian with analog quantum simulators?
Existing sampling-based quantum supremacy protocols heavily rely on certain statistical and computational properties of state ensembles formed by applying random unitaries to a fixed state~\cite{aaronson2011computational, bermejo2018architectures, bouland2019complexity}. 
The projected ensemble emerging from generic quantum dynamics may also possess the requisite properties and our work could lead to a new approach to quantum supremacy using analog quantum simulators.

\newpage
\noindent {\bf Acknowledgments.}\quad
We thank Adam Bouland, Fernando Brand\~ao, Aram Harrow, Wen Wei Ho, Nicholas Hunter-Jones, Anand Natarajan, and Hannes Pichler for valuable discussions.  This work was partly supported by the Institute for Quantum Information and Matter, an NSF Physics Frontiers Center (NSF Grant PHY- 1733907), the NSF CAREER award (1753386), the AFOSR YIP (FA9550-19-1-0044), the DARPA ONISQ program (W911NF2010021), the Army Research Office MURI program (W911NF2010136), and the NSF QLCI program (2016245).
JSC is supported by a Junior Fellowship from the Harvard Society of Fellows, as well as in part by
the Department of Energy under grant {DE}-{SC0007870}.
HH is supported by the J.~Yang \& Family Foundation.
FH is supported by the Fannie \& John Hertz Foundation.
JC acknowledges support from the IQIM postdoctoral fellowship.
ALS acknowledges support from the Eddleman Quantum graduate fellowship.
SC acknowledges support from the Miller Institute for Basic Research in Science.
\bibliography{refs}

\newpage

\onecolumngrid

\appendix

\section{Details of numerical simulations}
\label{sec:numericsdetails}

\subsection{$k$-th Moment of the Haar Ensemble}

To quantify the degree of randomness of our projected ensembles, we computed the trace distance $\Delta^{(k)}$ between the $k$-th moments of the projected ensemble and the Haar ensemble:
\begin{equation}
    \Delta^{(k)} = \frac{1}{2} \left\| \rho_\mathcal{E}^{(k)} - \rho_\textrm{Haar}^{(k)} \right\|_1~.
\end{equation}
Here $\rho_\textrm{Haar}^{(k)}$ is the $k$-th moment averaged over the Haar ensemble. For a Hilbert space $\mathcal{H}$ with dimension $d$, it has the form~\cite{harrow2013church}:
\begin{align}
\label{E:kthmoment1}
\rho_\textrm{Haar}^{(k)} &= \EX_{\Psi \sim \text{Haar}(d)}\!\left[(|\Psi\rangle \langle \Psi|)^{\otimes k}\right]\\
&= \frac{\sum_{\pi \in S_k} \text{Perm}_{\mathcal{H}^{\otimes k}}(\pi)}{d(d+1)\cdots(d+k-1)}\,.
\end{align}

Here, $S_k$ is the symmetric group on $k$ elements, and $\pi \in S_k$ is an element of the group.  $\text{Perm}_{\mathcal{H}^{\otimes k}}$ is a representation of $S_k$ on $\mathcal{H}^{\otimes k}$ which permutes the tensor factors according to
\begin{equation}
\text{Perm}_{\mathcal{H}^{\otimes k}}(\pi) \, |\Psi_1\rangle \otimes \cdots \otimes |\Psi_k\rangle = |\Psi_{\pi^{-1}(1)}\rangle \otimes \cdots \otimes |\Psi_{\pi^{-1}(k)}\rangle\,.
\end{equation}
The inverses in the subscripts are chosen so that $\text{Perm}_{\mathcal{H}^{\otimes k}}(\pi) \cdot \text{Perm}_{\mathcal{H}^{\otimes k}}(\pi') = \text{Perm}_{\mathcal{H}^{\otimes k}}(\pi \circ \pi')$ (i.e., the representation is a homomorphism of $S_k$, and not an antihomomorphism).  It is readily checked that Eq.~\eqref{E:kthmoment1} can be written as
\begin{align}
\frac{\sum_{\pi \in S_k}\text{Perm}_{\mathcal{H}^{\otimes k}}(\pi)}{d(d+1)\cdots(d+k-1)} = \frac{\Pi_k}{\binom{d+k-1}{k}}
\end{align}
where $\Pi_k$ is simply the projector onto the symmetric subspace of $\mathcal{H}^{\otimes k}$ (that is, the invariant subspace under $S_k$); this subspace has dimension $\binom{d+k-1}{k}$.

\subsection{Finite sampling error from the empirical Haar ensemble}

In the main text, we constructed projected ensembles of size $2^{N_B}$, with $N_B$ the size of the complement subsystem. In Figs.~2d, 3b, and 4b, we compared the system size scaling of $\Delta^{(k)}$ against the trace distance $\Delta^{(k)}_\text{em}$ of the \textit{empirical Haar ensemble}, an ensemble formed by sampling from the Haar ensemble $2^{N_B}$ times. In particular, we have
\begin{align}
    \rho_\text{em}^{(k)} &= \frac{1}{2^{N_B}} \sum_{i=1}^{2^{N_B}} \left( \vert \Psi_i \rangle \langle \Psi_i \vert \right)^{\otimes k}~,~\quad\vert \Psi_i \rangle \sim \text{Haar}(d)\\
    \Delta^{(k)}_\text{em}  &= \EX_{\Psi_1, \Psi_2,... \sim \text{Haar}(d)}\!\left[\frac{1}{2} \lVert\rho^{(k)}_\text{em} - \rho^{(k)}_\text{Haar}\rVert_1 \right]
\end{align}
This comparison is made to estimate the degree to which the error $\Delta^{(k)}$ in our projected ensemble is due to its finite size. We estimate $\Delta^{(k)}_\text{em}$ for various values of $N_B$ and find that $\Delta^{(k)}_\text{em}$ scales exponentially as $1/\sqrt{2^{N_B}}$.

\section{Additional numerical simulations of ergodic Hamiltonian models}
\label{sec:morenumerics}

\subsection{Random coupling model}

In the main text we presented results for the ergodic QIMF model. Here we discuss an additional ergodic model: a spin-1/2 model with random all-to-all interactions,
\begin{equation}
    H = \sum_{\substack{i,j=1\\i<j}}^N \sum_{\substack{\mu,\nu \in \{x,y,z\} \\ (\mu,\nu) \neq (z,z) }} J_{ij}^{\mu, \nu} \sigma^{\mu}_i \sigma^{\nu}_j~,
    \label{eq:rand_coup}
\end{equation}
where $J_{ij}^{\mu, \nu}$ are random variables drawn from i.i.d.~normal distributions: $J_{ij}^{\mu, \nu}\sim N(0, 1/N)$. Such random coupling models are paradigmatic examples of quantum chaotic systems. The variance of the couplings is chosen so that $\text{Tr}(H^2)/2^N \sim O(N)$.
Our model does not have $\sigma^z_i \sigma^z_j$ terms so that our initial state $|\Psi_0 \rangle = | 0 \rangle^{\otimes N}$ has zero energy, and accordingly is regarded to be at infinite temperature.

\begin{figure*}[t!]
	\centering
	\includegraphics[width=1\textwidth]{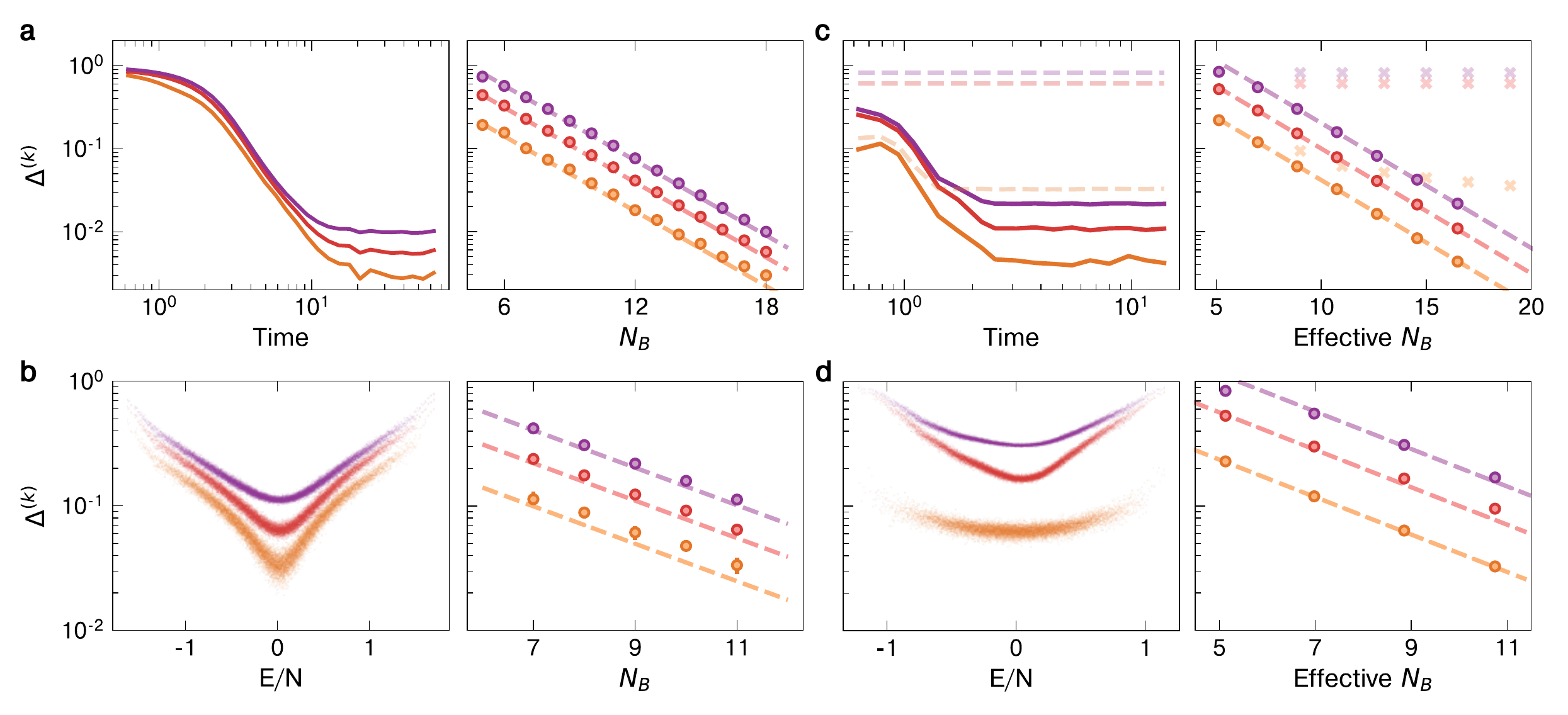}
	\caption{
	\textbf{Emergent quantum state designs in random coupling and random hopping models.}
	\textbf{a,}
	Time evolution and \textbf{b,} eigenstate data for the random all-to-all coupling model in Eq.~\eqref{eq:rand_coup}. The trace distances $\Delta^{(k)}$ for $k=1,2$ and $3$ are plotted in orange, red and purple respectively, for $N_A=3$ subsystems (\textbf{a}, left). Time evolution data for $N=21$ is plotted, along with the system size scaling of the late-time $\Delta^{(k)}$ for $N=8$ to $21$ (\textbf{a}, right). We also plot $\Delta^{(k)}$ for each eigenstate for $N=14$ (\textbf{b}, left), along with the system size scaling of the eigenstate data near $E=0$ for $N=10$ to $14$ (\textbf{b}, right).
	\textbf{c,} Time evolution and \textbf{d,} eigenstate data for the random hopping model in Eq.~\eqref{eq:rand_hop}. Data is presented for $N_A=5$ subsystems, in the $S^z_\text{tot.} = 0$ sector. In \textbf{c}, time evolution data for $N=24$ is plotted along with system size scaling for $N=10,12,..., 24$. To illustrate the failure of projecting onto all strings $z_B$, we plot the results of this na\"ive procedure with $N_A=3$ in light colors and dashes. In \textbf{d}, eigenstate data for $N=16$ is plotted, along with its system size scaling for $N=12,14,16$ and $18$. System size scaling is plotted against the effective $N_B$, which is $\log_2$ of the number of strings $z_B$ which is post-selected. In each plot, the error bars are smaller than the marker sizes.} \vspace{-0.5cm}
	\label{Fig5}
\end{figure*}

As shown in Fig.~5a, we see excellent convergence to $k$-designs for our projected ensembles constructed from time-evolved quenched states as well as eigenstates. The late-time $\Delta^{(k)}$ values are very close to $\Delta^{(k)}_\text{em}$. The average $\Delta^{(k)}$ values for infinite-temperature eigenstates are also close to $\Delta^{(k)}_\text{em}$, and show clear exponential decay (Fig.~5b). Notably, we do not average over disorder realizations: each data point and time series are computed with fixed disorder realizations. This provides additional support for our conjecture that ergodic Hamiltonian systems generate approximate $k$-designs via time-evolved states and eigenstates at infinite temperature.

\vspace{1em}
\subsection{Random hopping model with U(1) symmetry}

Having established that chaotic models such as the QIMF and the all-to-all random coupling model provide projected ensembles which converge to $k$-designs, we next ask whether quantum models with symmetries do so as well. We specifically study a random hopping model
\begin{equation}
    H = \sum_{\substack{i,j=1\\i<j}}^N J_{ij}^{+} \left(\sigma^x_i \sigma^x_j + \sigma^y_i\sigma^y_j \right) +  J_{ij}^{-} \left(\sigma^x_i \sigma^y_j - \sigma^y_i\sigma^x_j \right)~
    \label{eq:rand_hop}
\end{equation}
where $J_{ij}^{\pm}$ are random variables drawn from i.i.d.~normal distributions: $J_{ij}^{\pm}\sim N(0, 1/N)$. Like the all-to-all random coupling model, this model is also expected to exhibit chaos. However, the specific interaction terms are chosen such that this model has a $U(1)$ symmetry: there is conservation of the total magnetization $S^z_\text{tot.} = \sum_j S^z_j$. The model can equivalently be viewed as describing hard-core bosons with random complex all-to-all hopping amplitudes.

Any bitstring in the $z$-basis is a suitable infinite-temperature quench state. For $N$ even, we choose the initial state $|\Psi_0\rangle = |0101  \cdots 01 \rangle$ which lies in the $S^z_\text{tot.} = 0$ sector. Under time evolution by our chaotic Hamiltonian, the wavefunction is ergodic in the $S^z_\text{tot.} = 0$ Hilbert space.

Na\"ively forming the projected ensemble, we do not find convergence to an approximate $k$-design. This is shown in the light curves in Fig.~5c. With increasing system size, $\Delta^{(k)}$ for $k=1,2,3$ saturate at a non-zero value. This is because of large correlations between the bitstrings $z_B$ and the projected state $|\Psi_A(z_B)\rangle$: if $z_B$ has total magnetization $s_B$, $|\Psi_A(z_B)\rangle$ necessarily has total magnetization $S^z_\text{tot.}-s_B$ due to the global conservation law. The Hilbert space $\mathcal{H}_A$ naturally decomposes into a direct sum of multiple sectors enumerated by the magnetization $s_A$. Accordingly, the projected ensemble now produces multiple ensembles, one for each sector. 

In order to properly account for the $U(1)$ conservation law, instead of projecting onto all strings $z_B$, we post-select a subset of all strings: bitstrings with fixed total magnetization, e.g.~$s_B = 1/2$. The projected states $|\Psi_A(z_B)\rangle$ will also have fixed magnetization, and we then ask whether this ensemble forms a $k$-design in the subspace of $\mathcal{H}_A$ with magnetization $s_A = S^z_\text{tot.}-s_B$. In our numerical examples, we present results with $N_A = 5$ qubits and $s_A = -1/2, s_B = 1/2$. The relevant subspace of $\mathcal{H}_A$ has dimension $10$, far smaller than $2^{N_A} = 32$.

Fig.~5c,d show the results of our symmetry resolution. We now see excellent convergence towards a $k$-design. As with the random coupling model, the late-time $\Delta^{(k)}$ values are very close to $\Delta^{(k)}_\textrm{em}$. In order to make a fair comparison, we plot the late-time $\Delta^{(k)}$ against an ``effective $N_B$", defined as $\log_2$ of the number of post-selected strings $z_B$. $\Delta^{(k)}_\text{em}$ is also computed by sampling the same number of times.

Using this post-selection procedure, we find excellent $k$-design convergence for projected ensembles from eigenstates near $E=0$. Our results indicate that our basic approach remains valid for chaotic models with additional symmetries if these symmetries are properly addressed.

\section{Proof of main theorems}
\label{sec:proofsthm}

\subsection{Proof sketch for the main theorems}

We begin with a sketch of the proof of Theorem~\ref{Thm:HaarThm}.  We will use streamlined notation for clarity.
Recall that for a generator state $|\Phi\rangle$ on $\mathcal{H}$, the projected ensemble $\mathcal{E}_{\Phi,A} = \{p_z\,,\,|\Phi_z\rangle\}$ has
\begin{align}
p_z &= \langle \Phi | P_z |\Phi\rangle \\
|\Phi_z\rangle &= \big(\mathds{1}_A \otimes \langle z|_B\big)|\Phi\rangle/\sqrt{p_z}\,. \label{eq:phiz-def}
\end{align}
It will be convenient to define the unnormalized state
\begin{equation}
|\widetilde{\Phi}_z\rangle := \big(\mathds{1}_A \otimes \langle z|_B\big) |\Phi\rangle \quad \text{for}\quad z \in \{0,1\}^{N_B}
\end{equation}
so that $p_z = \langle \widetilde{\Phi}_z| \widetilde{\Phi}_z\rangle$ and $|\Phi_z\rangle = |\widetilde{\Phi}_z\rangle/\sqrt{p_z}$.  We note that in our new notation, we can write
\begin{align}
&\EX_{\Psi \sim \mathcal{E}_{\Phi, A}}\!\left[\big(|\Psi\rangle\langle\Psi|\big)^{\otimes k}\right]\\
&= \sum_{z \in \{0,1\}^{N_B}} p_z \,\big(|\Phi_z\rangle \langle \Phi_z|\big)^{\otimes k}\\
&= \sum_{z \in \{0,1\}^{N_B}} \frac{\big(|\widetilde{\Phi}_z\rangle \langle \widetilde{\Phi}_z|\big)^{\otimes k}}{\langle \widetilde{\Phi}_z | \widetilde{\Phi}_z\rangle^{k-1}} \equiv A(\ket{\Phi})\,.
\end{align}
The key to establishing Theorem~\ref{Thm:HaarThm} is understanding the random tensor $A(\ket{\Phi})$ when $\ket{\Phi}$ is a Haar-random state. Here, the Haar-random state is a normalized vector chosen uniformly at random in $d = d_A d_B$ complex dimensions, where $d = 2^N, d_A = 2^{N_A}, d_B = 2^{N_B}$ are the dimensions of the Hilbert space with $N$, $N_A$, and $N_B$ qubits respectively. Two ingredients are needed to understand $A(\ket{\Phi})$:
\begin{enumerate}
\item The expectation value of $A(\ket{\Phi})$ over the Haar ensemble.
\item The concentration of $A(\ket{\Phi})$ around its expectation. This will tell us that with high probability, $A(\ket{\Phi})$ is close to its expectation.
\end{enumerate}
Let us discuss these two ingredients in more detail.

Using the fact that the uniform measure (i.e., the Haar measure) on the complex sphere is invariant under any unitary rotations, we can show that $p_z$ and $|\Phi_z\rangle \langle \Phi_z|$ are independent random variables.
Furthermore, $|\Phi_z\rangle \langle \Phi_z|$ is a uniform random vector from a $d_A$-dimensional complex sphere.
Hence, we have
\begin{equation}
\EX_{\Phi\sim\mathrm{Haar}(d)} A(\ket{\Phi}) = \EX_{\Psi\sim\mathrm{Haar}(d_A)} \big(|\Psi \rangle \langle \Psi |\big)^{\otimes k}
\end{equation}
which is the $k$-th moment of the uniform ensemble.
This means that the expectation of the $k$-th moment of the projected ensemble of a randomly selected many-body wavefunction $\ket{\Phi}$ reproduces the $k$-th moment of the Haar ensemble in $d_A$ complex dimensions.
Hence the expectation $\EX_{\Phi\sim\mathrm{Haar}(d)} A(\ket{\Phi})$ is exactly equal to the desired quantity.

To understand the concentration of $A(\ket{\Phi})$ around its expectation, we recall the well-known result that the uniform distribution over a high-dimensional sphere has a very sharp concentration.
For any Lipschitz function on the sphere, the fluctuation around its expectation value is small and the probability of having a large fluctuation decays exponentially.
This is known as Levy's lemma and allows us to upper bound the probability that $A(\ket{\Phi})$ is far from its expectation.
Together with the expectation identity, we have
\begin{align}
&\underset{\Phi \sim \text{Haar}(d)}{\text{Prob}}\!\left[ \, \left\| A(\ket{\Phi}) - \underset{\Psi \sim \text{Haar}(d_A)}{\EX}\left[(|\Psi\rangle \langle \Psi|)^{\otimes k}\right]\right\|_1 \geq \varepsilon \right] \leq 2 d_A^{2k} \, \exp\left(- \frac{d_B \,\varepsilon^2}{18 \pi^3 (2k - 1) \,d_A^{4k}} \right)\,.
\end{align}
The asymptotic relation in Theorem~\ref{Thm:HaarThm} follows immediately from the above probabilistic statement. The detailed proof of Theorem~\ref{Thm:HaarThm} is given in Appendix~\ref{prf:thm1}.

\begin{figure}[t!]
	\centering
	\includegraphics[width=0.748\columnwidth]{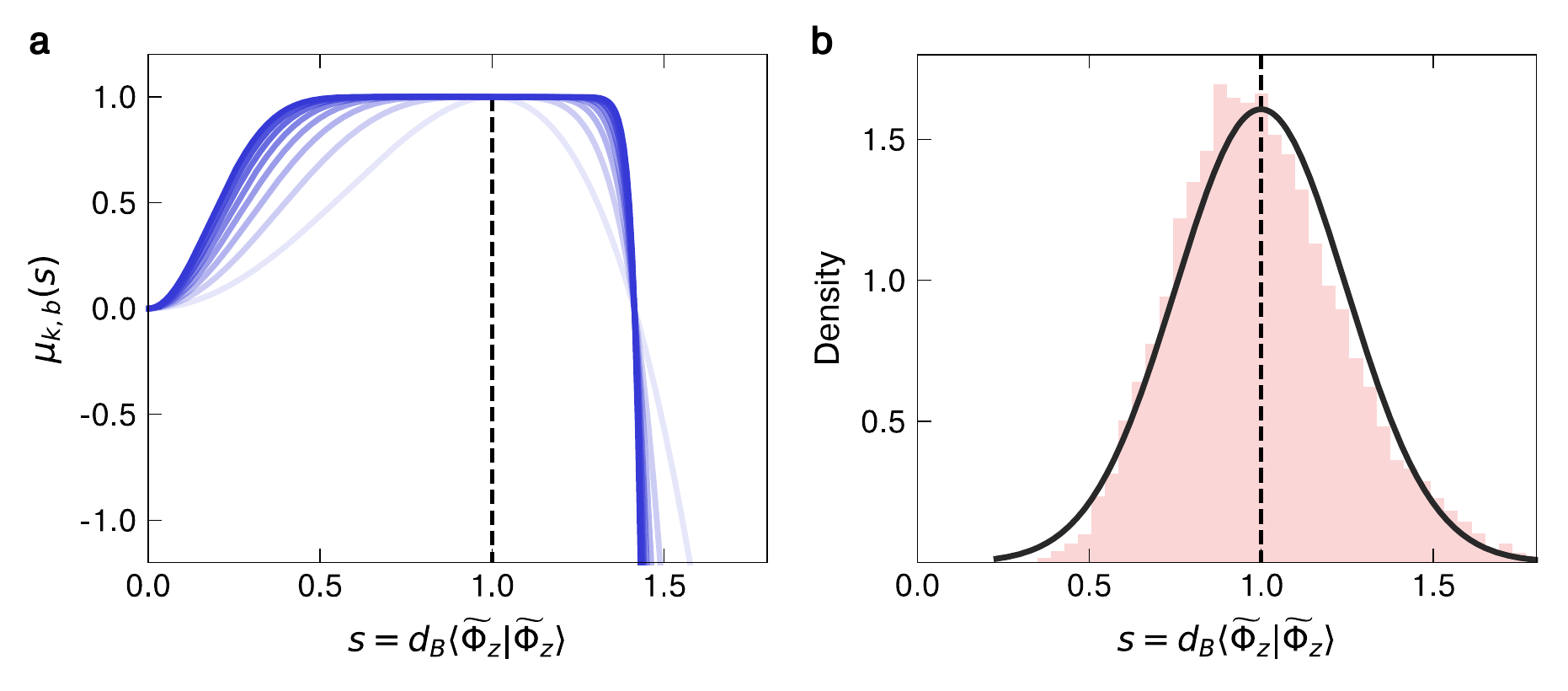}
	\vspace{-0.4cm}
	\caption{
	\textbf{Visualization of the modulating function and the concentration of $d_B \langle \widetilde{\Phi}_z | \widetilde{\Phi}_z\rangle$.}
	\textbf{a,} The modulating function $\mu_{k, b}(s)$ defined in Eq.~\eqref{eq:modulatingmu} for varying values of $b$ from $2$ to $16$. Darker colors correspond to higher $b$. We fix $k = 2$ in the plot.
	\textbf{b,} The concentration of $s = d_B \langle \widetilde{\Phi}_z | \widetilde{\Phi}_z\rangle$ for the Hilbert space dimension of subsystem $A$, i.e., $d_A = 2^{N_A}= 16$, when $|\Phi\rangle$ is sampled uniformly at random from the complex sphere. We sample $10000$ different vectors $|\Phi\rangle$ to generate the histogram. The black solid line is a kernel density estimation fit of the histogram. If $d_A$ is larger, then $s$ is more concentrated.
	} \vspace{-0.1cm}
	\label{FigMP}
\end{figure}

The proof of Theorem~\ref{Thm:tdesigngen} is very different from Theorem~\ref{Thm:HaarThm}.
First of all, the expectation of $A(\ket{\Phi})$ in Theorem~\ref{Thm:HaarThm} is computed using the invariance property of the Haar measure, which does not hold for a measure that only forms a state design.
Furthermore, Levy's lemma only holds for the Haar measure, so we cannot resort to Levy's lemma to control its statistical fluctuations.  To prove Theorem~\ref{Thm:tdesigngen}, we make use of a technique used in the context of solving linear systems on a quantum computer~\cite{childs2017quantum}.
The key idea is to add a modulating function that approximates $A(\ket{\Phi})$ by a polynomial function in $\ket{\Phi}$.  In particular,
\begin{align}
A(\ket{\Phi}) &= \sum_{z \in \{0,1\}^{N_B}} \frac{\big(|\widetilde{\Phi}_z\rangle \langle \widetilde{\Phi}_z|\big)^{\otimes k}}{\langle \widetilde{\Phi}_z | \widetilde{\Phi}_z\rangle^{k-1}} \\
\approx B(\ket{\Phi}) &= \sum_{z \in \{0,1\}^{N_B}} \frac{\big(|\widetilde{\Phi}_z\rangle \langle \widetilde{\Phi}_z|\big)^{\otimes k}}{\langle \widetilde{\Phi}_z | \widetilde{\Phi}_z\rangle^{k-1}} \mu_{k, b}(d_B \langle \widetilde{\Phi}_z | \widetilde{\Phi}_z\rangle ),
\end{align}
where the modulating function is given by
\begin{equation} \label{eq:modulatingmu}
\mu_{k, b}(s) = (1 - (1 - s^{2(k-1)})^b) \quad \mathrm{for }\,\,b\,\,\mathrm{ even}.
\end{equation}
Using the binomial expansion, we can check that $B(\ket{\Phi})$ is a polynomial function.
Furthermore, $\mu_{k, b}(s)$ is very close to one when $s = d_B \langle \widetilde{\Phi}_z | \widetilde{\Phi}_z\rangle$ is around one.
Taking $b$ larger allows us to better approximate the constant function $\mu_{k,b}(s) = 1$, which corresponds to the target expression $A(\ket{\Phi})$.
A visualization of $\mu_{k, b}(s)$ can be found in Fig.~\ref{FigMP}a.
If $|\Phi\rangle$ is sampled from the uniform measure on the quantum state space, then $s = d_B \langle \widetilde{\Phi}_z | \widetilde{\Phi}_z\rangle$ will concentrate around one; see Fig.~\ref{FigMP}b.
So for $|\Phi\rangle$ sampled from the uniform measure, we can show that $A(\ket{\Phi}) \approx B(\ket{\Phi})$.
However, $|\Phi\rangle$ is sampled from a quantum state design, so there is no guarantee that $s = d_B \langle \widetilde{\Phi}_z | \widetilde{\Phi}_z\rangle$ should concentrate around one.
To address this, we utilize the following result proved in Lemma~\ref{lem:defResidual}
\begin{equation}
\norm{A(\ket{\Phi}) - B(\ket{\Phi})}_1 \leq R(\ket{\Phi}),
\end{equation}
where $R(\ket{\Phi})$ is a polynomial function given by
\begin{equation}
\sum_{z \in \{0, 1\}^{N_B}} \langle \widetilde{\Phi}_z | \widetilde{\Phi}_z \rangle \left( 1 - \mu_{k, b}\left(d_B \langle \widetilde{\Phi}_z | \widetilde{\Phi}_z \rangle\right)\right).
\end{equation}
Hence, the approximation error can be upper bounded by a polynomial function that we have better control over.

After introducing the key quantities, the proof proceeds by bounding the error between (i) $A(\ket{\Phi})$ where $\ket{\Phi}$ is sampled from a state design, and (ii) the expectation of $A(\ket{\Psi})$ where $\ket{\Psi}$ is from the Haar measure. We use $B(\ket{\Phi})$ as an intermediate point of comparison:
\begin{align}
&\EX_{\Phi \sim (\varepsilon', k')\,\mathrm{design}} \norm{A(\ket{\Phi}) - \EX_{\Psi \sim \mathrm{Haar}} A(\ket{\Psi})}_1\\
&\leq \EX_{(\varepsilon', k')\,\mathrm{design}} \norm{A(\ket{\Phi}) - B(\ket{\Phi})}_1 + \EX_{(\varepsilon', k')\,\mathrm{design}} \norm{B(\ket{\Phi}) - \EX_{\mathrm{Haar}} A(\ket{\Psi})}_1\,.
\end{align}
The first term can be upper bounded by $R(\ket{\Phi})$ and we can use the fact that state designs approximate the expectation of any polynomial function under the Haar measure.  Accordingly,
\begin{align}
&\EX_{(\varepsilon', k')\,\mathrm{design}} \norm{A(\ket{\Phi}) - B(\ket{\Phi})}_1 \leq \EX_{(\varepsilon', k')\,\mathrm{design}} R(\ket{\Phi}) \approx \EX_{\mathrm{Haar}} R(\ket{\Phi}).
\end{align}
In the second term, we can apply a similar idea by upper bounding the 1-norm with the 2-norm and utilizing the fact that $B(\ket{\Phi}) - \EX_{\mathrm{Haar}} A(\ket{\Psi})$ is a polynomial function in $\ket{\Phi}$. This gives
\begin{align}
\EX_{(\varepsilon', k')\,\mathrm{design}} \norm{B(\ket{\Phi}) - \EX_{\mathrm{Haar}} A(\ket{\Psi})}_1
&\leq \sqrt{ d_A^k \EX_{(\varepsilon', k')\,\mathrm{design}} \norm{B(\ket{\Phi}) - \EX_{\mathrm{Haar}} A(\ket{\Psi})}^2_2}\\
&\approx \sqrt{ d_A^k \EX_{\mathrm{Haar}} \norm{B(\ket{\Phi}) - \EX_{\mathrm{Haar}} A(\ket{\Psi})}^2_2}\,.
\end{align}
We then approximate $B(\ket{\Phi})$ by $A(\ket{\Phi})$ in the above expression, where the error can be upper bounded by $R(\ket{\Phi})^2$.
This gives the approximate relation
\begin{align}
\EX_{(\varepsilon', k')\,\mathrm{design}} \norm{B(\ket{\Phi}) - \EX_{\mathrm{Haar}} A(\ket{\Psi})}_1
&\lesssim \sqrt{ d_A^k \EX_{\mathrm{Haar}} \left(R(\ket{\Phi})^2 + \norm{A(\ket{\Phi}) - \EX_{\mathrm{Haar}} A(\ket{\Psi})}^2_2\right)}\,.
\end{align}
Using these steps, all our expressions contain only expectations over the Haar measure. More precisely,
\begin{align}
&\EX_{\Phi \sim (\varepsilon', k')\,\mathrm{design}} \norm{A(\ket{\Phi}) - \EX_{\Psi \sim \mathrm{Haar}} A(\ket{\Psi})}_1\\
&\lesssim  \EX_{\mathrm{Haar}} R(\ket{\Phi}) + \sqrt{ d_A^k \EX_{\mathrm{Haar}} R(\ket{\Phi})^2} + \sqrt{d_A^k \EX_{\mathrm{Haar}} \norm{A(\ket{\Phi}) - \EX_{\mathrm{Haar}} A(\ket{\Psi})}^2_2 }\,.
\end{align}
The first two terms are small because
\begin{enumerate}
\item $\mu_{k, b}(s)$ is close to one when $s$ is close to one.
\item If $|\Phi\rangle$ is sampled from the Haar measure, then $s = d_B \langle \widetilde{\Phi}_z | \widetilde{\Phi}_z\rangle$ will concentrate around one.
\end{enumerate}
See Figure~\ref{FigMP} for visualizations.
The third term is small due to Theorem~\ref{Thm:HaarThm}.
Finally, we can resort to Markov's inequality to show that with a probability of at least $0.99$, we have that $\norm{A(\ket{\Phi}) - \EX_{\Psi \sim \mathrm{Haar}} A(\ket{\Psi})}_1$ is small.
The full proof of Theorem~\ref{Thm:tdesigngen} is given in Appendix~\ref{prf:thm2}.

\subsection{Useful identities}

We collect here several useful identities that we will leverage throughout the remainder of the Appendix.  The first identity is
\begin{equation}
\label{E:tildedHaar1}
\EX_{\Phi \sim \text{Haar}(d)}\!\left[ (|\widetilde{\Phi}_z\rangle \langle \widetilde{\Phi}_z|)^{\otimes k}\right] = \frac{\sum_{\pi \in S_k} \text{Perm}_{\mathcal{H}_A^{\otimes k}}(\pi) }{d(d+1)\cdots (d+k-1)}\,.
\end{equation}
Taking the trace, we find
\begin{equation}
\label{E:tildedHaar2}
\EX_{\Phi \sim \text{Haar}(d)}\!\left[ \langle \widetilde{\Phi}_z|\widetilde{\Phi}_z\rangle^{k}\right] = \frac{d_A(d_A+1)\cdots (d_A + k-1)}{d(d+1)\cdots (d+k-1)}\,.
\end{equation}
The related identity
\begin{equation}
\label{E:tildedHaar3}
\EX_{\Phi \sim \text{Haar}(d)}\!\left[|\langle \widetilde{\Phi}_z|\widetilde{\Phi}_y\rangle|^{2k}\right] = \frac{k!\,d_A(d_A+1)\cdots (d_A + k-1)}{d(d+1)\cdots (d+2k-1)}\qquad \text{for }x \not = y
\end{equation}
will likewise be useful.

\subsection{Proof of Theorem~\ref{Thm:HaarThm}}
\label{prf:thm1}

The main ingredient to prove Theorem~\ref{Thm:HaarThm} is Levy's lemma.
Levy's lemma states that if a function is Lipschitz continuous, then it will concentrate around it's expectation value if the variable is sampled uniformly from a high-dimensional complex sphere.

\begin{lemma}[Levy's Lemma] \label{lem:levy}
Let $f : \mathbb{S}^{2d-1} \to \mathbb{R}$ satisfying $|f(v) - f(w)| \leq \eta \|v-w\|_2$\,.  Then for any $\delta \geq 0$, we have
\begin{equation}
\text{\rm Prob}_{\Phi \sim \text{\rm Haar}(d)}\!\left[\,\left|f(\Phi) - \EX_{\Psi \sim \text{\rm Haar}(d)}\!\left[f(\Psi)\right]\,\right| \geq \delta\right] \leq 2 \exp\left(- \frac{2 d\, \delta^2}{9 \pi^3 \eta^2}\right)\,.
\end{equation}
\end{lemma}

Before applying Levy's lemma, we will need to define a function of interest. In particular, we will consider individual entries in the main quantity $\sum_{z \in \{0,1\}^{N_B}} p_z \,\big(|\Phi_z\rangle \langle \Phi_z|\big)^{\otimes k}$.
Let $\{|i\rangle\}_{i \in \{0,1\}^{tN_A}}$ be the standard basis on $\mathcal{H}_A^{\otimes k}$, and write $|i\rangle = \bigotimes_{k=1}^t |i^{(k)}\rangle$ where each $|i^{(k)}\rangle \in \mathcal{H}_A$.  Similarly write $|j\rangle = \bigotimes_{k=1}^t |j^{(k)}\rangle$.  Consider the function $f_{ij} : \mathbb{S}^{2d-1} \to \mathbb{R}$ defined by
\begin{align}
\label{E:fijdef1}
f_{ij}(|\Phi\rangle) &= \langle i| \left(\sum_{z \in \{0,1\}^{N_B}} \frac{\big(|\widetilde{\Phi}_z\rangle \langle \widetilde{\Phi}_z|\big)^{\otimes k}}{\langle \widetilde{\Phi}_z | \widetilde{\Phi}_z \rangle^{k-1}}\right) |j\rangle \\
&= \sum_{z \in \{0,1\}^{N_B}} \frac{\prod_{\ell=1}^k \langle i^{(\ell)}|\widetilde{\Phi}_z\rangle \langle \widetilde{\Phi}_z|j^{(\ell)}\rangle}{\langle \widetilde{\Phi}_z | \widetilde{\Phi}_z \rangle^{k-1}}\,.
\end{align}
A nice property of this function is that it is Lipschitz continuous and hence Levy's lemma can be applied.

\begin{lemma}[Lipschitz constant] \label{lem:lipsconst}
We have
\begin{equation}
\left\| \frac{d}{d|\Phi\rangle} \, f_{ij}(|\Phi\rangle) \right\|_2 \leq 2 (2k - 1) = \eta.
\end{equation}
\end{lemma}
\begin{proof}
Since our $f_{ij}$ is differentiable, we can choose any $\eta$ such that
\begin{equation}
\eta \geq \left\| \frac{d}{d|\Phi\rangle} \, f_{ij}(|\Phi\rangle) \right\|_2\,.
\end{equation}
We have
\begin{align}
\label{E:tobound1}
\left\| \frac{d}{d|\Phi\rangle} \, f_{ij}(|\Phi\rangle) \right\|_2 &= \Bigg\| \sum_{\ell'=1}^k \sum_{z \in \{0,1\}^{N_B}} \frac{\prod_{\ell \not = \ell'} \langle i^{(\ell)}|\widetilde{\Phi}_z\rangle \langle \widetilde{\Phi}_z|j^{(\ell)}\rangle}{\langle \widetilde{\Phi}_z | \widetilde{\Phi}_z \rangle^{k-1}} \frac{d}{d|\Phi\rangle} \langle i^{(\ell')} | \widetilde{\Phi}_z\rangle \langle\widetilde{\Phi}_z| j^{(\ell')}\rangle \nonumber \\
& \qquad \qquad \qquad - (k-1) \sum_{z \in \{0,1\}^{N_B}} \frac{\prod_{\ell=1}^k \langle i^{(\ell)}|\widetilde{\Phi}_z\rangle \langle \widetilde{\Phi}_z|j^{(\ell)}\rangle}{\langle \widetilde{\Phi}_z | \widetilde{\Phi}_z \rangle^{k}} \, \frac{d}{d|\Phi\rangle} \langle \widetilde{\Phi}_z|\widetilde{\Phi}_z\rangle \Bigg\|_2 \nonumber \\
&\leq \Bigg\| \sum_{\ell'=1}^k \sum_{z \in \{0,1\}^{N_B}} \frac{\prod_{\ell \not = \ell'} \langle i^{(\ell)}|\widetilde{\Phi}_z\rangle \langle \widetilde{\Phi}_z|j^{(\ell)}\rangle}{\langle \widetilde{\Phi}_z | \widetilde{\Phi}_z \rangle^{k-1}} \frac{d}{d|\Phi\rangle} \langle i^{(\ell')} | \widetilde{\Phi}_z\rangle \langle\widetilde{\Phi}_z| j^{(\ell')}\rangle \Bigg\|_2 \nonumber \\
& \qquad \qquad \qquad \qquad + 2(k-1) \Bigg\|\sum_{z \in \{0,1\}^{N_B}} \frac{\prod_{\ell=1}^k \langle i^{(\ell)}|\widetilde{\Phi}_z\rangle \langle \widetilde{\Phi}_z|j^{(\ell)}\rangle}{\langle \widetilde{\Phi}_z | \widetilde{\Phi}_z \rangle^{k}} \, P_z |\Phi\rangle \Bigg\|_2 \nonumber \\
\end{align}
where we have simply explicitly evaluated the derivatives and used the triangle inequality for the 2-norm.  Let us bound each of the term terms on the right-hand side in turn.  For the first one, we have
\begin{align}
&\Bigg\| \sum_{\ell'=1}^k \sum_{z \in \{0,1\}^{N_B}} \frac{\prod_{\ell \not = \ell'} \langle i^{(\ell)}|\widetilde{\Phi}_z\rangle \langle \widetilde{\Phi}_z|j^{(\ell)}\rangle}{\langle \widetilde{\Phi}_z | \widetilde{\Phi}_z \rangle^{k-1}} \frac{d}{d|\Phi\rangle} \langle i^{(\ell')} | \widetilde{\Phi}_z\rangle \langle\widetilde{\Phi}_z| j^{(\ell')}\rangle \Bigg\|_2 \nonumber \\
& \qquad \qquad \qquad = \Bigg\| \sum_{\ell'=1}^k \sum_{z \in \{0,1\}^{N_B}} \frac{\prod_{\ell \not = \ell'} \langle i^{(\ell)}|\widetilde{\Phi}_z\rangle \langle \widetilde{\Phi}_z|j^{(\ell)}\rangle}{\langle \widetilde{\Phi}_z | \widetilde{\Phi}_z \rangle^{k-1}} \left( \left( |j^{(\ell')}\rangle \langle i^{(\ell')}| + |i^{(\ell')}\rangle \langle j^{(\ell')}| \right) \otimes |z\rangle \langle z|_B \right)|\Phi\rangle \Bigg\|_2 \,.
\end{align}
Writing $b_{\ell',x} = \frac{\prod_{\ell \not = \ell'} \langle i^{(\ell)}|\widetilde{\Phi}_z\rangle \langle \widetilde{\Phi}_z|j^{(\ell)}\rangle}{\langle \widetilde{\Phi}_z | \widetilde{\Phi}_z \rangle^{k-1}}$
and $M_{\ell'} = |j^{(\ell')}\rangle \langle i^{(\ell')}| + |i^{(\ell')}\rangle \langle j^{(\ell')}|$ (which is Hermitian), the above evaluates to
\begin{align}
\label{E:firsttermsimplified1}
&\left(\sum_{\ell',p'=1}^k \sum_{z \in \{0,1\}^{N_B}}  b_{p',x}^* b_{\ell',x} \langle \Phi |\left( M_{p'}^\dagger M_{\ell'}  \otimes |z\rangle \langle z|_B \right) |\Phi\rangle \right)^{1/2} \nonumber \\
&\qquad \qquad \qquad \qquad \qquad \qquad \qquad \leq  \left(\sum_{\ell', p'=1}^k \sum_{z \in \{0,1\}^{N_B}}  |b_{p',x}^*|\, |b_{\ell',x}|\, \langle \Phi |\left( |M_{p'}^\dagger M_{\ell'}|  \otimes |z\rangle \langle z|_B \right) |\Phi\rangle \right)^{1/2}
\end{align}
where $|A| := \sqrt{A^\dagger A}$.  But since
\begin{align}
|b_{\ell',x}| &= \left|\text{tr}\!\left\{\left(\bigotimes_{\ell \not = \ell'} |j^{(\ell)}\rangle \langle i^{(\ell)}|\right) \cdot \left( \frac{\big(| \widetilde{\Phi}_z\rangle \langle \widetilde{\Phi}_z|\big)^{\otimes (k-1)}}{\langle \widetilde{\Phi}_z| \widetilde{\Phi}_z\rangle^{k-1}}\right)\right\} \right|\nonumber \leq \left\| \bigotimes_{\ell \not = \ell'} |j^{(\ell)}\rangle \langle i^{(\ell)}|\right\|_2 \, \left\| \frac{\big(| \widetilde{\Phi}_z\rangle \langle \widetilde{\Phi}_z|\big)^{\otimes (k-1)}}{\langle \widetilde{\Phi}_z| \widetilde{\Phi}_z\rangle^{k-1}} \right\|_2 = 1\,,
\end{align}
Eq.~\eqref{E:firsttermsimplified1} is less than or equal to
\begin{align}
\left(\sum_{\ell',p'=1}^k \sum_{z \in \{0,1\}^{N_B}}  \langle \Phi |\left( |M_{p'}^\dagger M_{\ell'}|  \otimes |z\rangle \langle z|_B \right) |\Phi\rangle \right)^{1/2} &= \left(\sum_{\ell',p'=1}^t \langle \Phi |\left( |M_{p'}^\dagger M_{\ell'}|  \otimes \mathds{1}_B \right) |\Phi\rangle \right)^{1/2} \nonumber \\
&\leq \left(\sum_{\ell',p'=1}^t \| M_{p'}^\dagger M_{\ell'}\|_\infty \right)^{1/2}\,.
\end{align}
Since $\| M_{p'}^\dagger M_{\ell'}\|_\infty \leq 4$, the above is $\leq 2t$.

Now we bound the term in the last line of Eq.~\eqref{E:tobound1}.  Letting $c_z = \frac{\prod_{\ell=1}^k \langle i^{(\ell)}|\widetilde{\Phi}_z\rangle \langle \widetilde{\Phi}_z|j^{(\ell)}\rangle}{\langle \widetilde{\Phi}_z | \widetilde{\Phi}_z \rangle^{k}}$, the term can be written as
\begin{align}
\label{E:secondtermsimplified1}
2(k-1) \left(\sum_{z \in \{0,1\}^{N_B}} |c_z|^2 \langle \Phi | P_z |\Phi\rangle \right)^{1/2}\,.
\end{align}
Using the bound
\begin{align}
|c_z| &= \left|\text{tr}\!\left\{\left(\bigotimes_{\ell = 1}^k |j^{(\ell)}\rangle \langle i^{(\ell)}|\right) \cdot \left( \frac{\big(| \widetilde{\Phi}_z\rangle \langle \widetilde{\Phi}_z|\big)^{\otimes k}}{\langle \widetilde{\Phi}_z| \widetilde{\Phi}_z\rangle^k}\right)\right\} \right|\nonumber \leq \left\| \bigotimes_{\ell =1}^k |j^{(\ell)}\rangle \langle i^{(\ell)}|\right\|_2 \, \left\| \frac{\big(| \widetilde{\Phi}_z\rangle \langle \widetilde{\Phi}_z|\big)^{\otimes k}}{\langle \widetilde{\Phi}_z| \widetilde{\Phi}_z\rangle^k} \right\|_2 = 1\,,
\end{align}
we can upper bound Eq.~\eqref{E:secondtermsimplified1} by
\begin{equation}
2(k-1) \left(\sum_{z \in \{0,1\}^{N_B}} \langle \Phi | P_z |\Phi\rangle \right)^{1/2} = 2(k-1)\,.
\end{equation}

Putting our bounds together, we have
\begin{align}
\left\| \frac{d}{d|\Phi\rangle} \, f_{ij}(|\Phi\rangle) \right\|_2 \leq 2t + 2(k-1) = 2(2k-1)\,,
\end{align}
which is the desired result.
\end{proof}

While the above bound on the Lipschitz constant, along with Levy's lemma, guarantee that each function $f_{ij}$ concentrates around the its expectation value, the expectation value of $f_{ij}$ has not yet been computed.
The following lemma shows that the expectation value of $f_{ij}$ is closely connected to the quantum state $k$-design on subsystem $A$.

\begin{lemma}[Expectation value identity]\label{lemma:expvaliden} We have the identity
\begin{equation}
\EX_{\Phi \sim \mathrm{Haar}(d)}\!\left[
\sum_{z \in \{0,1\}^{N_B}} \frac{(|\widetilde{\Phi}_z\rangle \langle \widetilde{\Phi}_z|)^{\otimes k}}
{\braket{\widetilde{\Phi}_z|\widetilde{\Phi}_z}^{k-1}}\right]
=
\EX_{\Psi\sim \mathrm{Haar}(d_A)}\!\left[(|\Psi\rangle \langle \Psi|)^{\otimes k}\right]\,.
\end{equation}
Therefore the expectation of $f_{ij}$ is given by
\begin{equation}
\EX_{\Phi \sim \mathrm{Haar}(d)} \left[ f_{ij}(\ket{\Phi}) \right] = \bra{i} \EX_{\Psi\sim \mathrm{Haar}(d_A)}\!\left[(|\Psi\rangle \langle \Psi|)^{\otimes k}\right] \ket{j}
\end{equation}
\end{lemma}
\begin{proof}
Observe that
\begin{equation}
\frac{(|\widetilde{\Phi}_z\rangle \langle \widetilde{\Phi}_z|)^{\otimes k}}
{\braket{\widetilde{\Phi}_z|\widetilde{\Phi}_z}^{k-1}}
= (|\Phi_z\rangle \langle \Phi_z|)^{\otimes k}
\braket{\widetilde{\Phi}_z|\widetilde{\Phi}_z}.
\end{equation}
Furthermore, from a technical lemma given below, we have that $\ket{\Phi_z}$ is independent from
$\braket{\widetilde{\Phi}_z|\widetilde{\Phi}_z}$.
Using this fact and the linearity of expectation, we compute
\begin{equation}
\begin{split}
\EX_{\Phi\sim \mathrm{Haar}(d)}\!\left[
\sum_{z \in \{0,1\}^{N_B}} \frac{(|\widetilde{\Phi}_z\rangle \langle\widetilde{\Phi}_z|)^{\otimes k}}
{\braket{\widetilde{\Phi}_z|\widetilde{\Phi}_z}^{k-1}}\right]
&=
\sum_{z \in \{0,1\}^{N_B}}
\EX_{\Phi \sim\mathrm{Haar}(d)}\!\left[(|\Phi_z\rangle \langle \Phi_z|)^{\otimes k}\right]
\EX_{\Phi\sim\mathrm{Haar}(d)}\!\left[
\braket{\widetilde{\Phi}_z|\widetilde{\Phi}_z}\right] \\
&=
\EX_{\Psi\sim\mathrm{Haar}(d_A)}\!
\left[(|\Psi\rangle \langle \Psi|)^{\otimes k}\right]
\left(\sum_{z \in \{0,1\}^{N_B}} \frac{1}{d_B}\right)
\end{split}
\end{equation}
as desired.
\end{proof}

During the evaluation of the expectation value, we used a property that the normalized state $|\Phi_z\rangle \langle \Phi_z|$ and the corresponding probability $p_z$ are independent random variables. This is proven in the following lemma.

\begin{lemma}
\label{indep-lem}
If $\ket{\Phi}$ is a Haar-random state on
$\mathcal{H}$, then the random variables $|\Phi_z\rangle \langle \Phi_z|$ and $p_z$ are independent.
\end{lemma}
\begin{proof}
Define the map
$\mathcal{P}_z(|\Phi\rangle) := \braket{\Phi|P_z|\Phi}$, and let $\mathcal{R}_z$ be the map taking $\ket{\Phi}$ to the normalized
state $\ket{\Phi_z}$,
\begin{equation}
\mathcal{R}_z(|\Phi\rangle)
:= \text{tr}_B\!\left((\mathds{1}_A \otimes |z\rangle \langle z|_B) \cdot |\Phi\rangle \langle \Phi|\right) / \mathcal{P}_z(|\Phi\rangle)\,.
\end{equation}
Let $U_A$ be a Haar-random unitary operator on $\mathcal{H}_A$ and
let $U=U_A\otimes \mathds{1}_B$.  Because $U$ is unitary and
$\ket{\Phi}$ is Haar-random, the state $U\ket{\Phi}$ is also Haar-random.
Also note that
\begin{equation}
\mathcal{P}_z(U\ket{\Phi}) = \mathcal{P}_z(\ket{\Phi})\,, \qquad \mathcal{R}_z(U\ket{\Phi}) = U_A \mathcal{R}_z(\ket{\Phi}) U_A^\dagger
\end{equation}
where the equivalence is in the sense of random variables.  Therefore, for any functions $F$ and $G$,
\begin{equation}
\begin{split}
\EX_{\Phi\sim\textrm{Haar}(d)}\left[
F(\mathcal{R}_z(\ket{\Phi}))\,G(\mathcal{P}_z(\ket{\Phi})\right]
&=
\EX_{\substack{\Phi\sim\textrm{Haar}(d) \\ U_A \sim U(d_A)}}\left[
F(\mathcal{R}_z(U\ket{\Phi}))\,G(\mathcal{P}_z(\ket{\Phi}) \right]\\
&=
\EX_{\substack{\Phi\sim\textrm{Haar}(d) \\ U_A \sim U(d_A)}}\left[
F(U_A\mathcal{R}_z(\ket{\Phi})U_A^\dagger)
\,G(\mathcal{P}_z(\ket{\Phi})\right] \\
&=
\EX_{\Psi\sim\textrm{Haar}(d_A)}
\!\left[F(\ket{\Psi}\bra{\Psi})\right]
\EX_{\Phi\sim\textrm{Haar}(d)}\!\left[
G(\mathcal{P}_z(\ket{\Phi})\right]\,.
\end{split}
\end{equation}
To pass from the second line to the third, we used the fact that for
any state $\ket{\Psi}$ on $\mathcal{H}_A$, the state
$U_A\ket{\Psi}$ is an independent Haar-random state on $\mathcal{H}_A$.
This shows that $\mathcal{R}_z(\ket{\Phi})$ and
$\mathcal{P}_z(\ket{\Phi})$ are independent, as desired.
\end{proof}

With the above lemmas, we now proceed with the proof of Theorem~\ref{Thm:HaarThm}.
\begin{proof}[Proof of Theorem~\ref{Thm:HaarThm}]
Let us consider the function $f_{ij}(\ket{\Phi})$ defined in Eq.~\eqref{E:fijdef1}.
Because we have bound the Lipschitz constant $\eta \leq 4t - 2$, we can leverage Levy's lemma given in Lemma~\ref{lem:levy} to obtain
\begin{equation}
\text{Prob}_{\Phi \sim \text{Haar}(d)}\!\left[\left|f_{ij}(|\Phi\rangle) - \EX_{\Phi \sim \text{Haar}(d)}\!\left[f_{ij}(|\Phi\rangle)\right]\right| \geq \varepsilon \right] \leq 2 \, \exp\left(- \frac{d \,\varepsilon^2}{18 \pi^3 (2k - 1)} \right)\,.
\end{equation}
Performing a union bound and rescaling $\varepsilon \to \varepsilon/d_A^{2k}$, we have
\begin{equation}
\text{Prob}_{\Phi \sim \text{Haar}(d)}\!\left[\left|f_{ij}(|\Phi\rangle) - \EX_{\Phi \sim \text{Haar}(d)}\!\left[f_{ij}(|\Phi\rangle)\right]\right| \geq \frac{\varepsilon}{d_A^{2k}}\,,\,\text{for some}\,i,j \right] \leq 2 d_A^{2k} \, \exp\left(- \frac{d \,\varepsilon^2}{18 \pi^3 (2k - 1) \,d_A^{4k}} \right)\,.
\end{equation}
By comparing to the definition of $f_{ij}$ in Eq.~\eqref{E:fijdef1} and using Lemma~\ref{lemma:expvaliden} to obtain the expectation value of $f_{ij}$, the above is equivalent to the following concentration inequality
\begin{align}
&\text{Prob}_{\Phi \sim \text{Haar}(d)}\!\left[ \, \left\| \sum_{z \in \{0,1\}^{N_B}} \frac{\big(|\widetilde{\Phi}_z\rangle \langle \widetilde{\Phi}_z|\big)^{\otimes k}}{\langle \widetilde{\Phi}_z | \widetilde{\Phi}_z \rangle^{k-1}} - \EX_{\Phi \sim \text{Haar}(d)}\!\left[\sum_{z \in \{0,1\}^{N_B}} \frac{\big(|\widetilde{\Phi}_z\rangle \langle \widetilde{\Phi}_z|\big)^{\otimes k}}{\langle \widetilde{\Phi}_z | \widetilde{\Phi}_z \rangle^{k-1}}\right] \right\|_{\text{entrywise, }1} \geq \varepsilon \right] \nonumber \\
& \qquad \qquad \qquad \qquad \qquad \qquad \qquad \qquad \qquad \qquad \qquad \qquad \qquad \qquad \quad \leq 2 d_A^{2k} \, \exp\left(- \frac{d \,\varepsilon^2}{18 \pi^3 (2k - 1) \,d_A^{4k}} \right)
\end{align}
where $\|A\|_{\text{entrywise, }1} = \sum_{i,j}|A_{ij}|$.  Finally, using $\|A\|_{\text{entrywise, }1} \geq \|A\|_1$ and applying Lemma~\ref{lemma:expvaliden}, we find
\begin{align}
\label{E:1normconc1}
&\text{Prob}_{\Phi \sim \text{Haar}(d)}\!\left[ \, \left\| \sum_{z \in \{0,1\}^{N_B}} \frac{\big(|\widetilde{\Phi}_z\rangle \langle \widetilde{\Phi}_z|\big)^{\otimes k}}{\langle \widetilde{\Phi}_z | \widetilde{\Phi}_z \rangle^{k-1}} - \EX_{\Psi \sim \text{Haar}(d_A)}\left[(|\Psi\rangle \langle \Psi|)^{\otimes k}\right]\right\|_1 \geq \varepsilon \right] \leq 2 d_A^{2k} \, \exp\left(- \frac{d_B \,\varepsilon^2}{18 \pi^3 (2k - 1) \,d_A^{4k}} \right)\,.
\end{align}
We can see that if we have
\begin{equation}
d_B \geq \frac{18 \pi^3 (2k-1) d_A^{4k}}{\varepsilon^2}\left( 2t \log(d_A) + \log(2 / \delta) \right),
\end{equation}
then the ensemble $\mathcal{E}_{\Phi, A}$ forms an $\varepsilon$-approximate $t$-design with probability at least $1 - \delta$. Recall that $d_A = 2^{N_A}, d_B = 2^{N_B}$, hence taking a logarithm on both side of the above condition gives rise to
\begin{equation}
N_B = \Omega\left(k N_A + \log\left(\frac{1}{\varepsilon}\right) + \log \log\left(\frac{1}{\delta}\right)\right),
\end{equation}
which is the result stated in the main text.
\end{proof}

\subsection{Proof of Theorem~\ref{Thm:tdesigngen}}
\label{prf:thm2}

For convenience we restate Theorem~\ref{Thm:tdesigngen} in the following.

\begin{theorem}[Restatement of Theorem 2]
Let $\ket{\Psi}$ be a state chosen from an ensemble on $\mathcal{H}$ that forms an $\varepsilon'$-approximate $k'$-design. Then the projected ensemble $\mathcal{E}_{A, \Psi}$ forms an $\varepsilon$-approximate $k$-design with probability at least $1-\delta$ if
\begin{align}
N_B &= \Omega\left( k N_A + \log\left( \tfrac{1}{\varepsilon \delta} \right) \right),\\
k' &= \Omega\left( k \left(N_B +\log\left(\tfrac{1}{\varepsilon \delta}\right) \right)\right),\\
\log(\tfrac{1}{\varepsilon'}) &= \Omega\left( k N_B \left(N_B + \log\left(\tfrac{1}{\varepsilon \delta}\right)\right) \right),\\
N_A &= \Omega\left( \log(N_B) +\log(k) + \log\log\left(\tfrac{1}{\varepsilon \delta}\right) \right).
\end{align}
\end{theorem}

In the following subsections, we will begin with a discussion of a generalized Levy's lemma that provides sharper concentration for quadratic functions. We will then give a general structure of the proof and provide the detailed proof of several technical lemmas afterwards.

\subsubsection{Higher order concentration}
\label{sec:highordercon}

In the proof of Theorem~\ref{Thm:HaarThm}, Levy's lemma plays a crucial role in establishing the desired statement.
Levy's lemma tells us about a sharp concentration when the random variables are sampled from a high-dimensional sphere.
We will continue to utilize concentration on high-dimensional spheres for the proof of Theorem~\ref{Thm:tdesigngen}.
However, the original statement of Levy's lemma does not provide good bounds for Theorem~\ref{Thm:tdesigngen}.
We will instead make use of a higher order variant of Levy's lemma.
In this section we recall a higher order variant of Levy's lemma that was established in~\cite{bobkov2019higher}, and use it to provide a simple proof of a concentration inequality for the quantity $\braket{\pt_z|\pt_z}$.
This higher order concentration results provided in~\cite{bobkov2019higher} is a useful tool for obtaining concentration inequalities for polynomial functions on measures satisfying a log-Sobolev inequality, and may prove useful in other problems arising in quantum information theory.
For this work we only need the following result from~\cite{bobkov2019higher}:
\begin{theorem}[Theorem~\ref{Thm:HaarThm}.13 in \cite{bobkov2019higher}]
\label{sphere-quadratic}
Let $f$ be a $C^2$-smooth function on an open neighborhood
of the sphere $\Sphere^{N-1}$ with
$\int_{\Sphere^{N-1}} f \diff\sigma_{N-1}=0$.  If
\begin{equation}
\label{l2-grad-bd}
\int_{\Sphere^{N-1}} \|\nabla f\|_2^2\diff\sigma_{N-1} \leq \frac{1}{N}
\end{equation}
and $\|\Hess f(\theta)\|_{\infty} \leq 1$ for all
$\theta\in\Sphere^{N-1}$ then
\begin{equation}
\int_{\Sphere^{N-1}}
\exp((N-1)|f|/(8e)) \diff \sigma_{N-1} \leq 2.
\end{equation}
\end{theorem}

As a corollary, we obtain the following result for quadratic forms.
\begin{corollary}[Concentration for quadratic forms]
\label{quad-conc-bd}
Let $Q:\Real^N\to\Real^N$ be a linear map satisfying
$\|Q\|_{2} \leq \frac{1}{2}$ and $\|Q\|_{\infty}\leq \frac{1}{2}$, and let
\[
q_0 = \int_{\Sphere^{N-1}}\theta^TQ\theta\diff\sigma_{N-1}(\theta).
\]
Then
\begin{equation}
\label{quadratic-exp-bd}
\int_{\Sphere^{N-1}}
\exp((N-1)(\theta^TQ\theta-q_0)/(8e))\diff\sigma_{N-1}
\leq 2.
\end{equation}
\end{corollary}
\begin{proof}
Let $f(x) = x^TQx-a_0$.  Then $\nabla f(x) = 2Qx$, so
\begin{equation}
\int_{\Sphere^{N-1}} \|\nabla f\|_2^2 \diff\sigma_{N-1} =
4\int_{\Sphere^{N-1}} \theta^T Q^TQ\theta \diff\sigma_{N-1}(\theta).
\end{equation}
Let $\{u_j\}_{j=1}^N$ be an orthonormal basis of eigenvectors of
$Q^TQ$ with eigenvalues $\lambda_j^2$.  Then the latter integral is equal
to
\begin{equation}
4
\sum_j \lambda_j^2 \int_{\Sphere^{N-1}} (\theta\cdot u_j)^2
\diff \sigma_{N-1}(\theta)
= \frac{4}{N} \sum_j \lambda_j^2 = 4n^{-1}\|Q\|_{2}^2.
\end{equation}
If $\|Q\|_{2}\leq 1/2$ then $f$ satisfies the condition~\eqref{l2-grad-bd}.
Moreover $\Hess f = 2Q$, so $\|\Hess f\|_{\infty} = 2\|M\|_{\infty}$.  We can
therefore apply Theorem~\ref{sphere-quadratic} to $f$ to
obtain~\eqref{quadratic-exp-bd}.
\end{proof}

We only need to use Corollary~\eqref{quad-conc-bd} in the case that $Q$ is an
orthogonal projection.
\begin{corollary}[Concentration for quadratic forms with projectors]
\label{proj-lem}
Let $V\subset \Real^N$ be a subspace of $\Real^N$ with dimension $m$,
and let $P_V$ be the orthogonal projection onto $V$.  Then
\begin{equation}
\label{proj-conc-bd}
\EX_{\theta \sim \Sphere^{N-1}}\!\left[
\exp\left( \frac{(N-1)|\|P_V\theta\|_2^2-m/N|}{8e\sqrt{2m}}\right)\right] \leq 2\,,,
\end{equation}
and in particular, for $\theta$ sampled uniformly from the sphere,
\begin{equation}
\label{markov-bd}
\text{\rm Prob}_{\theta \sim \mathbb{S}^{N-1}}\!\left[|\|P_V\theta\|^2 - m/N| \geq \delta\right]
\leq 2 \exp\left(-\frac{1}{8e\sqrt{2}}\,(N-1)m^{-1/2}\delta\right).
\end{equation}
\end{corollary}
\begin{proof}
Observe that $\|P_V\|_{2}^2 = m$ and $\|P_V\|_{\infty}=1$, so the operator
$(2m)^{-1/2}P_V$ satisfies the conditions of Corollary~\ref{quad-conc-bd},
from which~\eqref{proj-conc-bd} follows.
Then~\eqref{markov-bd} follows from Markov's inequality applied
to~\eqref{proj-conc-bd}.
\end{proof}

Specializing to the case that $\ket{\Phi}$ is a Haar-random
quantum state on $\mathcal{H}$ and $P_V$ is the
projector $P_z = \mathds{1}_A\otimes \ket{x}\bra{x}$, we deduce an
exponential concentration inequality for $\braket{\pt_z|\pt_z}$.
\begin{corollary}[Concentration for probability in a projected ensemble]\label{cor:secondmain} We have the bound
\begin{equation}
\label{psix-conc-bd}
\text{\rm Prob}_{\Phi \sim \text{\rm Haar}(d)}\!\left[
\left|\langle \pt_z|\pt_z\rangle-\frac{1}{d_B}\right|\geq\delta\right]
\leq 2\exp\left(-\frac{1}{8e\sqrt{2}}\, d_A^{1/2}\,d_B\,\delta\right)\,.
\end{equation}
\end{corollary}

\subsubsection{General structure of the proof}

The proof is based on the idea of modulating the target expression, which is a rational function, with a high-degree polynomial to form a polynomial function.
More precisely, we will be interested in the following two functions:
\begin{align}
A(\ket{\Phi}) &= \sum_{z \in \{0, 1\}^{N_B}} \frac{ | \widetilde{\Phi}_z \rangle\! \langle \widetilde{\Phi}_z |^{\otimes k} }{ \langle \widetilde{\Phi}_z | \widetilde{\Phi}_z \rangle^{k-1}},\\
B(\ket{\Phi}) &= \sum_{z \in \{0, 1\}^{N_B}} \frac{ | \widetilde{\Phi}_z \rangle\! \langle \widetilde{\Phi}_z |^{\otimes k} }{ \langle \widetilde{\Phi}_z | \widetilde{\Phi}_z \rangle^{k-1}} \left( 1 - \left(1 - \left( \frac{d_B}{r} \langle \widetilde{\Phi}_z | \widetilde{\Phi}_z \rangle \right)^{2(k-1)}\right)^b \right), \label{eq:Bdef}
\end{align}
where $r, b > 0$ are parameters for tuning the approximation of polynomial function $B(\ket{\Phi})$ to the target expression $A(\ket{\Phi})$, which is a rational function.
Because the $1$ is cancelled in the binomial expansion
\begin{equation}
1 - \left(1 - \left( \frac{d_B}{r} \langle \widetilde{\Phi}_z | \widetilde{\Phi}_z \rangle \right)^{2(k-1)}\right)^b = \sum_{p=1}^b {b \choose p} (-1)^p \left( \frac{d_B}{r} \langle \widetilde{\Phi}_z | \widetilde{\Phi}_z \rangle \right)^{2(k-1)p},
\end{equation}
it is not hard to see that $B(\ket{\Phi})$ is indeed a polynomial function in the real and imaginary parts of $\ket{\Phi}$.

We will consider $\ket{\Phi}$ to be sampled from an $(\varepsilon', k')$-design.
We will use the basic Markov inequality to bound the concentration. A higher order concentration inequality can also be used but would require $k'$ to be higher.
The central quantity in Markov inequality is the expectation value of the error
\begin{align}
\EX_{\Phi \sim (\varepsilon', k')\,\mathrm{design}} \norm{A(\ket{\Phi}) - \EX_{\Psi \sim \mathrm{Haar}} A(\ket{\Psi})}_1.
\end{align}
We will use $B(\ket{\Phi})$ as a surrogate to obtain an upper bound on the above quantity. This is because $B(\ket{\Phi})$ is a polynomial function rather than a rational function in the real and imaginary parts of $\ket{\Phi}$.
A triangle inequality gives the following bound
\begin{align}
&\EX_{\Phi \sim (\varepsilon', k')\,\mathrm{design}} \norm{A(\ket{\Phi}) - \EX_{\Psi \sim \mathrm{Haar}} A(\ket{\Psi})}_1\\
&\leq \EX_{(\varepsilon', k')\,\mathrm{design}} \norm{A(\ket{\Phi}) - B(\ket{\Phi})}_1 + \EX_{(\varepsilon', k')\,\mathrm{design}} \norm{B(\ket{\Phi}) - \EX_{\Psi \sim \mathrm{Haar}} A(\ket{\Psi})}_1. \label{eq:twoterms}
\end{align}
We can now analyze the two terms independently by using properties of quantum designs.
For the first term, we will prove in Lemma~\ref{lem:defResidual} that the following inequality holds.
\begin{align} \label{def:Rfunc}
\norm{A(\ket{\Phi}) - B(\ket{\Phi})}_1
\leq \sum_{z \in \{0, 1\}^{N_B}} \langle \widetilde{\Phi}_z | \widetilde{\Phi}_z \rangle \left(1 - \left( \frac{d_B}{r} \langle \widetilde{\Phi}_z | \widetilde{\Phi}_z \rangle \right)^{2(k-1)}\right)^b.
\end{align}
Therefore we will define a polynomial function
\begin{equation}
\label{E:Rfunc1}
R(\ket{\Phi}) = \sum_{z \in \{0, 1\}^{N_B}} \langle \widetilde{\Phi}_z | \widetilde{\Phi}_z \rangle \left(1 - \left( \frac{d_B}{r} \langle \widetilde{\Phi}_z | \widetilde{\Phi}_z \rangle \right)^{2(k-1)}\right)^b,
\end{equation}
which is an upper bound on the approximation error between $A(\ket{\Phi})$ and $B(\ket{\Phi})$. Because a quantum design approximates any polynomial function, we have the following upper bound
\begin{align}
\EX_{(\varepsilon', k')\,\mathrm{design}} \norm{A(\ket{\Phi}) - B(\ket{\Phi})}_1 &\leq \EX_{(\varepsilon', k')\,\mathrm{design}} R(\ket{\Phi})\\
&\leq \EX_{\mathrm{Haar}} R(\ket{\Phi}) + \varepsilon' \times \mathrm{E}_1, \label{eq:boundfirstterm}
\end{align}
where the exact expression of $\mathrm{E}_1$ is given in Lemma~\ref{lem:Error1}. We will also apply a similar philosophy for bounding the second term in Eq.~\eqref{eq:twoterms} by first upper bounding the term by a polynomial function that turns expectations over designs into expectations over the Haar measure:
\begin{align}
\EX_{(\varepsilon', k')\,\mathrm{design}} \norm{B(\ket{\Phi}) - \EX_{\Psi \sim \mathrm{Haar}} A(\ket{\Psi})}_1 &\leq \EX_{(\varepsilon', k')\,\mathrm{design}} \sqrt{d_A^k} \norm{B(\ket{\Phi}) - \EX_{\Psi \sim \mathrm{Haar}} A(\ket{\Psi})}_2 \\
&\leq \sqrt{\EX_{(\varepsilon', k')\,\mathrm{design}} d_A^k \norm{B(\ket{\Phi}) - \EX_{\Psi \sim \mathrm{Haar}} A(\ket{\Psi})}^2_2}\\
&\leq \sqrt{\EX_{\mathrm{Haar}} d_A^k \norm{B(\ket{\Phi}) - \EX_{\Psi \sim \mathrm{Haar}} A(\ket{\Psi})}^2_2 + \varepsilon' \times \mathrm{E}_2}. \label{eq:boundsecondterm}
\end{align}
The first line follows from the relation between $1$-norm and $2$-norm.
The second line applies Jensen's inequality.
The third line follows from the observation that $d_A^k \norm{B(\ket{\Phi}) - \EX_{\Psi \sim \mathrm{Haar}} A(\ket{\Psi})}^2_2$ is a polynomial function in $\ket{\Phi}$. Therefore, we can turn the expectation over a quantum state design to one over the Haar measure, which incurs a small error of $\varepsilon' \times \mathrm{E}_2$. The exact expression of $\mathrm{E}_2$ is given in Lemma~\ref{lem:Error2}.
We will now upper bound $\EX_{\mathrm{Haar}} \norm{B(\ket{\Phi}) - \EX_{\Psi \sim \mathrm{Haar}} A(\ket{\Psi})}^2_2$ by turning $B(\ket{\Phi})$ back into $A(\ket{\Phi})$ and incurring an additional error:
\begin{align}
\EX_{\mathrm{Haar}} \norm{B(\ket{\Phi}) - \EX_{\Psi \sim \mathrm{Haar}} A(\ket{\Psi})}^2_2 &\leq 2 \EX_{\mathrm{Haar}} \norm{B(\ket{\Phi}) - A(\ket{\Phi})}^2_2 + 2 \EX_{\mathrm{Haar}} \norm{A(\ket{\Phi}) - \EX_{\Psi \sim \mathrm{Haar}} A(\ket{\Psi})}^2_2\\
&\leq 2 \EX_{\mathrm{Haar}} \norm{B(\ket{\Phi}) - A(\ket{\Phi})}^2_1 + 2 \EX_{\mathrm{Haar}} \norm{A(\ket{\Phi}) - \EX_{\Psi \sim \mathrm{Haar}} A(\ket{\Psi})}^2_2\\
&\leq 2 \EX_{\mathrm{Haar}} R(\ket{\Phi})^2 + 2 \EX_{\mathrm{Haar}} \norm{A(\ket{\Phi}) - \EX_{\Psi \sim \mathrm{Haar}} A(\ket{\Psi})}^2_2. \label{eq:insidesecondtermbound}
\end{align}
The first line follows from the triangle inequality and $(a + b)^2 \leq 2(a^2 + b^2), \forall a, b \in \mathbb{R}$. The second line follows from the relation between the $1$-norm and $2$-norm. The third line uses the inequality given in Eq.~\eqref{def:Rfunc}, which is proved in Lemma~\ref{lem:defResidual}.
We can now combine Eq.'s~\eqref{eq:twoterms},~\eqref{eq:boundfirstterm},~\eqref{eq:boundsecondterm},~and~\eqref{eq:insidesecondtermbound} to find
\begin{align}
&\EX_{\Phi \sim (\varepsilon', k')\,\mathrm{design}} \norm{A(\ket{\Phi}) - \EX_{\Psi \sim \mathrm{Haar}} A(\ket{\Psi})}_1\\
&\leq \EX_{\mathrm{Haar}} R(\ket{\Phi}) + \varepsilon' \times \mathrm{E}_1 + \sqrt{ 2 \EX_{\mathrm{Haar}} R(\ket{\Phi})^2 + 2 \EX_{\mathrm{Haar}} \norm{A(\ket{\Phi}) - \EX_{\Psi \sim \mathrm{Haar}} A(\ket{\Psi})}^2_2 + \varepsilon' \times \mathrm{E}_2}\,.
\end{align}
In Lemma~\ref{lem:Residualfun}, we give upper bounds for $\EX_{\mathrm{Haar}} R(\ket{\Phi})$ and $\EX_{\mathrm{Haar}} R(\ket{\Phi})^2$.
Lemma~\ref{lem:Residualfun} is the key to Theorem~\ref{Thm:tdesigngen}.
The goodness of fit of the polynomial approximation $B(\ket{\Phi})$ to the target expression $A(\ket{\Phi})$ is reflected in upper bound of $\EX_{\mathrm{Haar}} R(\ket{\Phi})$ and $\EX_{\mathrm{Haar}} R(\ket{\Phi})^2$.
By changing the polynomial approximation, one may likely obtain an improved statement of Theorem~\ref{Thm:tdesigngen}. We leave open the choice of the optimal polynomial approximation, and focus mainly on the simple and tractable polynomial approximation given in Eq.~\eqref{eq:Bdef}.
The upper bounds for $\EX_{\mathrm{Haar}} R(\ket{\Phi})$ and $\EX_{\mathrm{Haar}} R(\ket{\Phi})^2$ rely on the higher order variant of Levy's lemma given in Appendix~\ref{sec:highordercon}.
In Lemma~\ref{lem:AAapprox}, we obtain an upper bound on $\EX_{\mathrm{Haar}} \norm{A(\ket{\Phi}) - \EX_{\mathrm{Haar}} A(\ket{\Phi})}^2_2$.
Along with Lemma~\ref{lem:Error1} that bounds $\varepsilon' E_1$, Lemma~\ref{lem:Error2} that bounds $\varepsilon' E_2$, we can show that for any $\varepsilon > 0$ as long as we choose $r = 2$ and $b$ to be an even integer with
\begin{align}
b &= \Omega\left( N_B + \log(1 / \varepsilon) \right),\\
N_A &= \Omega\left( \log(N_B) + \log(k) + \log\log(1 / \varepsilon)\right),\\
k' &= \Omega\left( b k \right),\\
\log(1 / \varepsilon') &= \Omega\left( \log(1 / \varepsilon) + k (b N_B + \log k + N_A) \right),\\
N_B &= \Omega\left( k N_A + \log\left(\frac{1}{\varepsilon}\right) \right),
\end{align}
we can obtain the following upper bounds
\begin{align}
\varepsilon' \times \mathrm{E}_1 &\leq \varepsilon, \qquad \varepsilon' \times \mathrm{E}_2 \leq \varepsilon^2, \\
\EX_{\mathrm{Haar}} R(\ket{\Phi}) &\leq \varepsilon, \qquad \EX_{\mathrm{Haar}} R(\ket{\Phi})^2 \leq \varepsilon^2, \\
\EX_{\mathrm{Haar}} \norm{A(\ket{\Phi}) - \EX_{\Psi \sim \mathrm{Haar}} A(\ket{\Psi})}^2_2 &\leq \varepsilon^2.
\end{align}
Together with Markov's inequality, we have the following concentration result:
\begin{equation}
\Proba_{\Phi \sim (\varepsilon', k')\,\mathrm{design}}
\left[ \norm{A(\ket{\Phi}) - \EX_{\Psi \sim \mathrm{Haar}} A(\ket{\Psi})}_1 \geq \tilde{\varepsilon} \right] \leq \frac{ 2 \varepsilon + \sqrt{5 \varepsilon^2} }{\tilde{\varepsilon}} = 5\, \frac{\varepsilon}{\tilde{\varepsilon}}
\end{equation}
for any $\varepsilon, \tilde{\varepsilon} > 0$.
Using Lemma~\ref{lemma:expvaliden} established in the proof of Theorem~\ref{Thm:HaarThm}, we have
\begin{equation}
\EX_{\Psi \sim \mathrm{Haar}} A(\ket{\Psi}) = \EX_{\Psi\sim \mathrm{Haar}(d_A)}\!\left[(|\Psi\rangle \langle \Psi|)^{\otimes k}\right]\,.
\end{equation}
Therefore the projected ensemble of a randomly sample state $\ket{\Phi}$ forms an $(\varepsilon, k)$-design on the local subsystem with dimension $d_A = 2^{N_A}$ under probability at least $1-\delta$ as long as the following conditions hold:
\begin{align}
N_B &= \Omega\left( k N_A + \log\left( \frac{1}{\varepsilon \delta} \right) \right),\\
k' &= \Omega\left( k \left(N_B +\log\left(\frac{1}{\varepsilon \delta}\right) \right)\right),\\
\log(1 / \varepsilon') &= \Omega\left( k N_B \left(N_B + \log\left(\frac{1}{\varepsilon \delta}\right)\right) \right),\\
N_A &= \Omega\left( \log(N_B) +\log(k) + \log\log\left(\frac{1}{\varepsilon \delta}\right) \right).
\end{align}
This concludes the proof of Theorem~\ref{Thm:tdesigngen}.

\subsubsection{Technical lemmas}

\begin{lemma}[Error bound on polynomial approximation] \label{lem:defResidual}
If $b$ is even, then
$\norm{A(\ket{\Phi}) - B(\ket{\Phi})}_1 \leq R(\ket{\Phi})$.
\end{lemma}
\begin{proof}
Recall the following definitions:
\begin{align}
A(\ket{\Phi}) &= \sum_{z \in \{0, 1\}^{N_B}} \frac{ | \widetilde{\Phi}_z \rangle\! \langle \widetilde{\Phi}_z |^{\otimes k} }{ \langle \widetilde{\Phi}_z | \widetilde{\Phi}_z \rangle^{k-1}},\\
B(\ket{\Phi}) &= \sum_{z \in \{0, 1\}^{N_B}} \frac{ | \widetilde{\Phi}_z \rangle\! \langle \widetilde{\Phi}_z |^{\otimes k} }{ \langle \widetilde{\Phi}_z | \widetilde{\Phi}_z \rangle^{k-1}} \left( 1 - \left(1 - \left( \frac{d_B}{r} \langle \widetilde{\Phi}_z | \widetilde{\Phi}_z \rangle \right)^{2(k-1)}\right)^b \right),\\
R(\ket{\Phi}) &= \sum_{z \in \{0, 1\}^{N_B}} \langle \widetilde{\Phi}_z | \widetilde{\Phi}_z \rangle \left(1 - \left( \frac{d_B}{r} \langle \widetilde{\Phi}_z | \widetilde{\Phi}_z \rangle \right)^{2(k-1)}\right)^b.
\end{align}
We note the definition of trace norm $\norm{\cdot}_1$ for Hermitian matrices:
\begin{equation}
\norm{X}_1 = \sup_{O: \norm{O}_\infty \leq 1} \tr(O X).
\end{equation}
Hence, we have
\begin{align}
\norm{A(\ket{\Phi}) - B(\ket{\Phi})}_1 &= \sup_{O: \norm{O}_\infty \leq 1} \tr\left(O \sum_{z \in \{0, 1\}^{N_B}} \frac{ | \widetilde{\Phi}_z \rangle\! \langle \widetilde{\Phi}_z |^{\otimes k} }{ \langle \widetilde{\Phi}_z | \widetilde{\Phi}_z \rangle^{k-1}} \left(1 - \left( \frac{d_B}{r} \langle \widetilde{\Phi}_z | \widetilde{\Phi}_z \rangle \right)^{2(k-1)}\right)^b \right)\\
&= \sup_{O: \norm{O}_\infty \leq 1} \sum_{z \in \{0, 1\}^{N_B}} \frac{ \left(\langle \widetilde{\Phi}_z |^{\otimes k}\right) O  \left(| \widetilde{\Phi}_z \rangle^{\otimes k}\right)}{ \langle \widetilde{\Phi}_z | \widetilde{\Phi}_z \rangle^{k-1}} \left(1 - \left( \frac{d_B}{r} \langle \widetilde{\Phi}_z | \widetilde{\Phi}_z \rangle \right)^{2(k-1)}\right)^b\\
&\leq \sup_{O: \norm{O}_\infty \leq 1} \sum_{z \in \{0, 1\}^{N_B}} \frac{ \langle \widetilde{\Phi}_z | \widetilde{\Phi}_z \rangle^{t} }{ \langle \widetilde{\Phi}_z | \widetilde{\Phi}_z \rangle^{k-1}}
\left|\left(1 - \left( \frac{d_B}{r} \langle \widetilde{\Phi}_z | \widetilde{\Phi}_z \rangle \right)^{2(k-1)}\right)^b\right| \\
&\leq \sup_{O: \norm{O}_\infty \leq 1} \sum_{z \in \{0, 1\}^{N_B}} \langle \widetilde{\Phi}_z | \widetilde{\Phi}_z \rangle
\left(1 - \left( \frac{d_B}{r} \langle \widetilde{\Phi}_z | \widetilde{\Phi}_z \rangle \right)^{2(k-1)}\right)^b\\
&= R(\ket{\Phi}).
\end{align}
This concludes the proof.
\end{proof}

\begin{lemma}[Quantum state design on the error bound] \label{lem:Error1}
For $k' \geq 2b(k-1) + 1$, we have
\begin{equation}
\EX_{(\varepsilon', k')\,\mathrm{design}} R(\ket{\Phi}) \leq \EX_{\mathrm{Haar}} R(\ket{\Phi}) + \varepsilon' E_1,
\end{equation}
where the error term is given by
\begin{equation}
E_1 = \left(1 + \left(\frac{d_B}{r}\right)^{2(k-1)}\right)^b.
\end{equation}
In particular, if we choose $r = 2 \leq d_B$ and recall that $d_B = 2^{N_B}$, then for any $\varepsilon > 0$ as long as
\begin{align}
k' &= \Omega\left( b k \right),\\
\log(1 / \varepsilon') &= \Omega\left( \log(1 / \varepsilon) + b k N_B \right),
\end{align}
we have the following upper bound on the error term: $\varepsilon' E_1 \leq \varepsilon$.
\end{lemma}
\begin{proof}
Letting $M_{zz} = \mathds{1}_A \otimes |z\rangle \langle z|$, for $k' \geq \ell$ we have
\begin{align}
\label{E:Mzzineq1}
&\left|\sum_{z \in \{0,1\}^{N_B}} \EX_{(\varepsilon',k')\text{ design}} \langle \pt_z| \pt_z\rangle^\ell  - \sum_{z \in \{0,1\}^{N_B}} \EX_{\text{Haar}} \langle \pt_z| \pt_z\rangle^\ell \right| \nonumber \\
=\,\,& \left|\text{tr}\left(\left(\mathds{1}^{\otimes(k'-\ell)}\otimes\sum_{z \in \{0,1\}^{N_B}} M_{zz}^{\otimes \ell}\right) \cdot \left(\EX_{(\varepsilon',k')\text{ design}} (|\Phi\rangle \langle \Phi|)^{\otimes k'} - \EX_{\text{Haar}} (|\Phi\rangle \langle \Phi|)^{\otimes k'} \right)\right) \right| \nonumber \\
\leq\,\,& \left\|\mathds{1}^{\otimes(k'-\ell)}\otimes\sum_{z \in \{0,1\}^{N_B}} M_{zz}^{\otimes \ell}\right\|_\infty \, \left\|\EX_{(\varepsilon',k')\text{ design}} (|\Phi\rangle \langle \Phi|)^{\otimes k'} - \EX_{\text{Haar}} (|\Phi\rangle \langle \Phi|)^{\otimes k'} \right\|_1 \nonumber \\
\leq\,\,& \varepsilon'
\end{align}
where in the last inequality we have used $\left\|\mathds{1}^{\otimes(k'-\ell)}\otimes\sum_{z \in \{0,1\}^{N_B}} M_{zz}^{\otimes \ell}\right\|_\infty = 1$.  Upon examining Eq.~\eqref{E:Rfunc1}, we see that for $k' \geq 2b(k-1) + 1$ we can leverage~\eqref{E:Mzzineq1} to achieve
\begin{equation}
\EX_{(\varepsilon', k')\,\mathrm{design}} R(\ket{\Phi}) \leq \EX_{\mathrm{Haar}} R(\ket{\Phi}) + \varepsilon' \left(1 + \left(\frac{d_B}{r}\right)^{2(k-1)}\right)^b\,,
\end{equation}
which is the desired bound.
\end{proof}

\begin{lemma}[Quantum state design on the polynomial approximation] \label{lem:Error2} For $k' \geq 4b(k-1) + 2$,
\begin{align}
\EX_{(\varepsilon', k')\,\mathrm{design}} d_A^k \norm{B(\ket{\Phi}) - \EX_{\Psi \sim \mathrm{Haar}}\!\left[A(\ket{\Psi})\right]}^2_2 \leq \EX_{\mathrm{Haar}} d_A^k \norm{B(\ket{\Phi}) - \EX_{\Psi \sim \mathrm{Haar}}\!\left[A(\ket{\Psi})\right]}^2_2 \nonumber + \varepsilon' E_2,
\end{align}
where the error term is given by
\begin{equation}
E_2 = \left(2 \,k! \left[-1 + \left(1 + \left(\frac{d_B}{r}\right)^{2(k-1)} \right)^b\right] + d_A^k\left[-1 + \left(1 + \left(\frac{d_B}{r}\right)^{2(k-1)} \right)^b\right]^2\right)\,.
\end{equation}
In particular, if we choose $r = 2 \leq d_B$ and recall that $d_B = 2^{N_B}$, then for any $\varepsilon > 0$ as long as
\begin{align}
k' &= \Omega\left( b k \right),\\
\log(1 / \varepsilon') &= \Omega\left( \log(1 / \varepsilon) + k (b N_B + \log k + N_A) \right),
\end{align}
we have the following upper bound on the error term: $\varepsilon' E_2 \leq \varepsilon^2$.
\end{lemma}
\begin{proof}
Recall from Lemma~\ref{lemma:expvaliden} that
\begin{equation}
\EX_{\Psi \sim \mathrm{Haar}(d)}\!\left[A(\ket{\Psi})\right] = \EX_{\Phi \sim \text{Haar}(d_A)} (|\Phi\rangle \langle \Phi|)^{\otimes k} = \frac{\Pi_{A,k}}{\binom{d_A+k-1}{k}}
\end{equation}
where in the last line, $\Pi_{A,k}$ is the projector onto the symmetric subspace of $\mathcal{H}_A^{\otimes k}$.  Then we can write $\norm{B(\Phi) - \EX_{\Psi \sim \mathrm{Haar}}\!\left[A(\ket{\Psi})\right]}^2_2$ as
\begin{align}
\norm{B(\Phi)- \EX_{\Psi \sim \mathrm{Haar}}\!\left[A(\ket{\Psi})\right]}^2_2 &= \text{tr}(B(\Phi)^2) - \frac{2 \,\text{tr}(B(\Phi))}{\binom{d_A+k-1}{k}} + \frac{1}{\binom{d_A+k-1}{k}}\,.
\end{align}
Using~\eqref{E:Mzzineq1}, we have for $k' \geq 2b(k-1) + 1$ that
\begin{equation}
\label{E:tracebound1}
d_A^k \EX_{(\varepsilon', k')\,\mathrm{design}} \left(- \frac{2 \,\text{tr}(B(\Phi))}{\binom{d_A+k-1}{k}}\right) \leq d_A^k  \EX_{\mathrm{Haar}} \left(- \frac{2 \,\text{tr}(B(\Phi))}{\binom{d_A+k-1}{k}}\right) +  \varepsilon' \cdot 2 \,k! \left[-1 + \left(1 + \left(\frac{d_B}{r}\right)^{2(k-1)} \right)^b\right]\,.
\end{equation}

To obtain a similar bound on $d_A^k \EX_{(\varepsilon', k')\,\mathrm{design}} \text{tr}(B(\Phi)^2)$, we need the following generalization of the inequality in~\eqref{E:Mzzineq1}.  Letting $M_{yz} = \mathds{1}_A \otimes |y\rangle \langle z|$,  for $k' \geq 2(p+q)$ we have
\begin{align}
\label{E:Myzineq1}
&\left|\sum_{y,z \in \{0,1\}^{N_B}} \EX_{(\varepsilon',t')\text{ design}} \,|\langle \pt_z| \pt_y\rangle|^{2p}\langle \pt_z| \pt_z\rangle^q \langle \pt_y| \pt_y\rangle^q - \sum_{y,z\in \{0,1\}^{N_B}} \EX_{\text{Haar}} \,|\langle \pt_z| \pt_y\rangle|^{2p}\langle \pt_z| \pt_z\rangle^q \langle \pt_y| \pt_y\rangle^q \right| \nonumber \\
=\,\,& \left|\text{tr}\left(\left(\mathds{1}^{\otimes(k'-2(p+q))}\otimes\sum_{y,z \in \{0,1\}^{N_B}} M_{yz}^{\otimes p}\otimes M_{yz}^{\otimes p} \otimes M_{zz}^{\otimes q} \otimes M_{yy}^{\otimes q}\right) \cdot \left(\EX_{(\varepsilon',k')\text{ design}} (|\Phi\rangle \langle \Phi|)^{\otimes k'} - \EX_{\text{Haar}} (|\Phi\rangle \langle \Phi|)^{\otimes k'} \right)\right) \right| \nonumber \\
\leq\,\,& \left\|\mathds{1}^{\otimes(k'-2(p+q))}\otimes\sum_{x,y \in \{0,1\}^{N_B}} M_{yz}^{\otimes p}\otimes M_{yz}^{\otimes p} \otimes M_{zz}^{\otimes q} \otimes M_{yy}^{\otimes q}\right\|_\infty \, \left\|\EX_{(\varepsilon',k')\text{ design}} (|\Phi\rangle \langle \Phi|)^{\otimes k'} - \EX_{\text{Haar}} (|\Phi\rangle \langle \Phi|)^{\otimes k'} \right\|_1 \nonumber \\
\leq\,\,& \varepsilon'
\end{align}
where we have used that $\left\|\mathds{1}^{\otimes(k'-2(p+q))}\otimes\sum_{y,z \in \{0,1\}^{N_B}} M_{yz}^{\otimes p}\otimes M_{yz}^{\otimes p} \otimes M_{zz}^{\otimes q} \otimes M_{yy}^{\otimes q}\right\|_\infty = 1$.  Using~\eqref{E:Myzineq1}, for $k' \geq 2t + 2(2b-1)(k-1) = 4b(k-1) + 2$ we have
\begin{align}
\label{E:tracebound2}
d_A^k \EX_{(\varepsilon', k')\,\mathrm{design}} \text{tr}(B(\Phi)^2)\leq d_A^k  \EX_{\mathrm{Haar}} \text{tr}(B(\Phi)^2) +  \varepsilon' \cdot d_A^k\left[-1 + \left(1 + \left(\frac{d_B}{r}\right)^{2(k-1)} \right)^b\right]^2\,.
\end{align}
Putting together~\eqref{E:tracebound1} and~\eqref{E:tracebound2}, we arrive at
\begin{align}
&\EX_{(\varepsilon', k')\,\mathrm{design}} d_A^k \norm{B(\ket{\Phi}) - \EX_{\Psi \sim \mathrm{Haar}}\!\left[A(\ket{\Psi})\right]}^2_2 \leq \EX_{\mathrm{Haar}} d_A^k \norm{B(\ket{\Phi}) - \EX_{\Psi \sim \mathrm{Haar}}\!\left[A(\ket{\Psi})\right]}^2_2 \nonumber \\
& \qquad \qquad \qquad \qquad \quad + \varepsilon' \left(2 \,k! \left[-1 + \left(1 + \left(\frac{d_B}{r}\right)^{2(k-1)} \right)^b\right] + d_A^k\left[-1 + \left(1 + \left(\frac{d_B}{r}\right)^{2(k-1)} \right)^b\right]^2\right)
\end{align}
for $k' \geq 4b(k-1) + 2$ as claimed.
\end{proof}

\begin{lemma}[Moment bounds on error function] \label{lem:Residualfun}
For $t \geq 2, r = 1$, $b$ even, and $d_A^{1/4} \geq 8t - 6$,
\begin{align}
\EX_{\mathrm{Haar}} R(\ket{\Phi}) &\leq \frac{d_B}{2^{b}} + 2 d_B^{2b (k-1) + 2} \exp\left(-\frac{1}{8e\sqrt{2}}\, d_A^{1/4}\right),\\
\EX_{\mathrm{Haar}} R(\ket{\Phi})^2 &\leq \frac{d_B^2}{2^{2b}} + 2 d_B^{4b (k-1) + 3} \exp\left(-\frac{1}{8e\sqrt{2}}\, d_A^{1/4}\right).
\end{align}
In particular, if we fix an $\varepsilon > 0$ and recall that $d_A = 2^{N_A}, d_B = 2^{N_B}$, then as long as
\begin{align}
b &= \Omega\left( N_B + \log(1 / \varepsilon) \right),\\
N_A &= \Omega\left( \log(N_B) + \log(k) + \log\log(1 / \varepsilon)\right),
\end{align}
we have the following moment bounds on the error function
\begin{align}
\EX_{\mathrm{Haar}} R(\ket{\Phi}) &\leq \varepsilon,\\
\EX_{\mathrm{Haar}} R(\ket{\Phi})^2 &\leq \varepsilon^2.
\end{align}
\end{lemma}
\begin{proof}
Recall the following definitions with $r = 1$:
\begin{align}
R(\ket{\Phi}) &= \sum_{z \in \{0, 1\}^{N_B}} \langle \widetilde{\Phi}_z | \widetilde{\Phi}_z \rangle \left(1 - \left( d_B \langle \widetilde{\Phi}_z | \widetilde{\Phi}_z \rangle \right)^{2(k-1)}\right)^b.
\end{align}
We note that $\langle \widetilde{\Phi}_z | \widetilde{\Phi}_z \rangle \in [0, 1]$ and we define an associated function
\begin{equation}
\gamma(s) = s\left( 1 - \left( d_B s \right)^{2(k-1)} \right)^b, \,\,\forall s \in [0, 1].
\end{equation}
For $s \in [(1 - 1 / d_A^{1/4}) / d_B, (1 + 1 / d_A^{1/4}) / d_B]$, we can see that $\gamma(s) \geq 0$ because $b$ is even. We now proceed to upper bound $\gamma(s)$ in this domain.
We have the following bound based on the condition that $s \leq (1 + 1 / d_A^{1/4}) / d_B$,
\begin{equation}
\left( d_B s \right)^{2(k-1)} - 1 \leq \left( 1 + \frac{1}{d_A^{1/4}} \right)^{2(k-1)} - 1 \leq \frac{2(k-1) / d_A^{1/4}}{1 - (2t-3) / d_A^{1/4}} \leq \frac{4(k-1)}{d_A^{1/4}}.
\end{equation}
The first inequality follows from the monotonicity of $(d_B s)^{2(k-1)} - 1$. The second inequality follows from the fact that $(1+x)^n \leq 1 + nx / (1 - (n-1) x), \,\,\forall z \in [-1, 1 / (n-1)], n > 1$. The third inequality uses the condition that $d_A  \geq 8t - 6 \geq 2(2t - 3)$.
We can also obtain another bound using $s \geq (1 - 1 / d_A^{1/4}) / d_B$,
\begin{equation}
1 - \left( d_B s \right)^{2(k-1)} \leq \frac{2(k-1)}{d_A^{1/4}},
\end{equation}
which follows from the fact that $(1+x)^n \geq 1 + nx, \,\,\forall x \geq -1, n \geq 1$.
Together we have $\forall s \in [(1 - 1 / d_A^{1/4}) / d_B, (1 + 1 / d_A^{1/4}) / d_B]$, the function $\gamma(s)$ is bounded as follows:
\begin{equation} \label{eq:gammas-good}
|\gamma(s)| \leq \left|\left( d_B s \right)^{2(k-1)} - 1\right|^b \leq \left(\frac{4(k-1)}{d_A^{1/4}}\right)^b \leq \frac{1}{2^b}.
\end{equation}
The second inequality uses the fact that $d_A^{1/4} \geq 8t - 6 \geq 8(k-1)$.
If the variable $s$ is not within that domain, but is in $[0, 1]$, we have
\begin{equation} \label{eq:badgamma}
|\gamma(s)| \leq d_B^{2b (k-1)}.
\end{equation}
We consider the event $G$ such that
\begin{equation}
\langle \pt_z|\pt_z\rangle \in \left[\frac{1 - 1 / d_A^{1/4}}{d_B}, \frac{1 + 1 / d_A^{1/4}}{d_B}\right], \,\,\forall z \in \{0, 1\}^{N_B},
\end{equation}
which is equivalent to the event that
\begin{equation}
\left|\langle \pt_z|\pt_z\rangle-\frac{1}{d_B}\right| \leq \frac{1}{d_A^{1/4} d_B}, \,\,\forall z \in \{0, 1\}^{N_B}.
\end{equation}
We now utilize the concentration result given in Corollary~\ref{cor:secondmain},
\begin{equation}
\text{\rm Prob}_{\Phi \sim \text{\rm Haar}(d)}\!\left[
\left|\langle \pt_z|\pt_z\rangle-\frac{1}{d_B}\right|\geq\delta\right]
\leq 2\exp\left(-\frac{1}{8e\sqrt{2}}\, d_A^{1/2}\,d_B\,\delta\right),
\end{equation}
to derive the probability upper bound for the complement of the event $G$
\begin{equation} \label{eq:badeventprob}
\text{\rm Prob}_{\Phi \sim \text{\rm Haar}(d)}\!\left[ \, G \,\, \mbox{did not happen} \,\right] \leq 2 d_B \exp\left(-\frac{1}{8e\sqrt{2}}\, d_A^{1/4} \right),
\end{equation}
which is obtained by taking the union bound.
We can now proceed to upper bound the first and second moments of $R(\ket{\Phi})$ by noting that $R(\ket{\Phi}) = \sum_{z \in \{0, 1\}^{N_B}} \gamma(\langle \widetilde{\Phi}_z | \widetilde{\Phi}_z \rangle)$.
When the event $G$ happens, we can use Eq.~\eqref{eq:gammas-good} to obtain
\begin{equation}
|R(\ket{\Phi})| \leq \frac{d_B}{2^b}\,.
\end{equation}
If the event $G$ did not happen, then using Eq.~\eqref{eq:badgamma}, we have
\begin{equation}
|R(\ket{\Phi})| \leq d_B^{2b (k-1) + 1}.
\end{equation}
Together for both $m = 1, 2$, we have
\begin{align}
\EX_{\mathrm{Haar}} R(\ket{\Phi})^m &\leq \left(\frac{d_B}{2^b}\right)^m \text{\rm Prob}_{\Phi \sim \text{\rm Haar}(d)}\!\left[ \, G \,\, \mbox{happened} \,\right] \\
&+ d_B^{m(2b (k-1) + 1)} \text{\rm Prob}_{\Phi \sim \text{\rm Haar}(d)}\!\left[ \, G \,\, \mbox{did not happen} \,\right] \\
&\leq \frac{d_B^m}{2^{bm}} + 2 d_B^{m(2b (k-1) + 1) + 1} \exp\left(-\frac{1}{8e\sqrt{2}}\, d_A^{1/4}\right),
\end{align}
where the last inequality uses the concentration result given in Eq.~\eqref{eq:badeventprob}.
The asymptotic bounds can be obtained from the above result under suitable choices of the parameters given in the statement of this lemma.
\end{proof}

\begin{lemma}[Second moment bound for projected ensemble from Haar measure] \label{lem:AAapprox}
\begin{equation}
\EX_{\Phi \sim \mathrm{Haar}(d)}\!\left[\norm{A(\ket{\Phi}) - \EX_{\Psi \sim \mathrm{Haar}(d)}\!\left[A(\ket{\Psi})\right] \, }^2_2\right] \leq 36 \pi^3 (2k -1) (d_A^{6k} / d_B)\,.
\end{equation}
In particular, recalling that $d_A = 2^{N_A}, d_B = 2^{N_B}$, then for any $\varepsilon > 0$, as long as
\begin{equation}
N_B = \Omega\left( k N_A + \log\left(\frac{1}{\varepsilon}\right) \right),
\end{equation}
we have the upper bound
\begin{equation}
\EX_{\Phi \sim \mathrm{Haar}(d)}\!\left[\norm{A(\ket{\Phi}) - \EX_{\Psi \sim \mathrm{Haar}(d)}\!\left[A(\ket{\Psi})\right] \, }^2_2\right] \leq \varepsilon^2.
\end{equation}
\end{lemma}
\begin{proof}
Notice that Eq.~\eqref{E:1normconc1} implies
\begin{equation}
\text{Prob}_{\Phi \sim \text{Haar}(d)}\!\left[ \norm{A(\ket{\Phi}) - \EX_{\Psi \sim \mathrm{Haar}(d)}\!\left[A(\ket{\Psi})\right] \, }^2_2 \geq y\right] \leq 2 d_A^{2k} \exp\left(- \frac{d_B y}{18 \pi^3 (2k-1) d_A^{4k}} \right)\,.
\end{equation}
Then we can use the fact that $\EX[X] = \int_{0}^{\infty} dx \,\, \text{Prob}[X \geq x]$ for any positive random variable $X$ to obtain
\begin{align}
&\EX_{\Phi \sim \mathrm{Haar}(d)}\!\left[\norm{A(\ket{\Phi}) - \EX_{\Psi \sim \mathrm{Haar}(d)}\!\left[A(\ket{\Psi})\right] \, }^2_2\right]\\
&= \int_{-\infty}^\infty dy\, \text{Prob}_{\Phi \sim \text{Haar}(d)}\!\left[ \norm{A(\ket{\Phi}) - \EX_{\Psi \sim \mathrm{Haar}(d)}\!\left[A(\ket{\Psi})\right] \, }^2_2 \geq y\right] \\
&\leq \int_{0}^\infty dy\, 2 d_A^{2k} \exp\left(- \frac{d_B y}{18 \pi^3 (2k-1) d_A^{4k}} \right) \\
&\leq 36 \pi^3 (2k-1) \frac{d_A^{6k}}{d_B}\,,
\end{align}
which establishes the bound.
\end{proof}

\pagebreak

\end{document}